\documentclass[11pt]{article}
\usepackage{amsmath}
\usepackage{amsfonts}
\usepackage{epsfig}
\usepackage{times}
\newcommand{\be}{\begin{equation}}
\newcommand{\ee}{\end{equation}}
\newcommand{\ba}{\begin{array}}
\newcommand{\ea}{\end{array}}
\newcommand{\bpma}{\begin{pmatrix}}
\newcommand{\epma}{\end{pmatrix}}
\newcommand{\bea}{\begin{eqnarray}}
\newcommand{\eea}{\end{eqnarray}}

\newcommand{\rar}{\rightarrow}
\newcommand{\p}{\partial}
\newcommand{\ol}{\overline}
\newcommand{\ti}{\tilde}
\newcommand{\la}{\langle}
\newcommand{\ra}{\rangle}

\renewcommand{\l}{\newline\null}
\def\figskip{\vskip .5cm plus 3mm minus 2mm}
\def\hbar{h\!\!\!/}
\begin{document}
\begin{titlepage}
July 1997  \hfill PAR-LPTHE 97/32
\vskip 1.5cm
\begin{center}
{\bf 
AN ELECTROWEAK \boldmath{$SU(2)_L \times U(1)$}
GAUGE THEORY OF \boldmath{$J=0$} MESONS.
}
\end{center}
\vskip 3mm
\centerline{B. Machet
     \footnote[1]{Member of ``Centre National de la Recherche Scientifique''}
     \footnote[2]{E-mail: machet@lpthe.jussieu.fr}
     }
\vskip 3mm
\centerline{{\em Laboratoire de Physique Th\'eorique et Hautes Energies,}
     \footnote[3]{LPTHE tour 16\,/\,1$^{er}\!$ \'etage,
          Universit\'e P. et M. Curie, BP 126, 4 place Jussieu,
          F 75252 PARIS CEDEX 05 (France).}
}
\centerline{\em Universit\'es Pierre et Marie Curie (Paris 6) et Denis
Diderot (Paris 7);} \centerline{\em Unit\'e associ\'ee au CNRS URA 280.}
\vskip .7cm
{\parskip=2pt plus 2pt minus 1pt\fontsize{10}{11}\selectfont
\begin{center}
Extended version of the talk 
``Custodial symmetry in a $SU(2)_L \times U(1)$ gauge theory of $J=0$ mesons''
given at the $2^{nd}$ International Symposium on Symmetries 
in Subatomic Physics,\l
Seattle (Washington, USA), June 25th-29th 1997.
\end{center}
\vskip .5cm
{\bf Abstract:} 
I display  all $J=0$ scalar and pseudoscalar representations  of the standard
$SU(2)_L \times U(1)$ group of electroweak interactions which transform 
like sets of fermion-antifermion composite fields. They can fit into quadruplets
of definite $CP$ quantum numbers.  $SU(2)_L \times U(1)$ is embedded in a
natural way, compatible with the Glashow-Salam-Weinberg model for quarks,
into the chiral group $U(N)_L \times U(N)_R$, $N$ being the (even)
number of ``flavours''. It involves a unitary $N/2 \times N/2$ ``mixing
matrix'' for fermions  which are however only considered here as mathematical
objects in the fundamental representation of $U(N)$.

The electroweak gauge Lagrangian for the $J=0$ particles exhibits a chiral
$SU(2)_L \times SU(2)_R$ symmetry at the limit when the hypercharge coupling
$g'$ goes to zero. It is spontaneously broken down to its diagonal $SU(2)_V$
subgroup, which includes the electromagnetic $U(1)$, spanning a bridge to
an explanation of electric charge quantization.
Chiral are electroweak spontaneous breaking are identical. The consequences
for the nature of the Goldstone bosons are examined.

Comparison with recent works by Cho {\em et al.} unraveling dyon-like 
solution is an electroweak model with the same structure suggests that 
electric-magnetic duality may be realized here, with the occurrence of a 
strongly interacting sector.

$SU(2)_L \times U(1)$ allows one mass scale per quadruplet; as each
decomposes into one triplet and one singlet of $SU(2)_V$, the custodial
symmetry doubles the amount of masses allowed for the $2N^2$ mesons,
from  $N^2/2$  to $N^2$.
They bear no special connection with the nature of their ``fermionic
content'' and share with the leptonic sector the same arbitrariness.
No information on the mass of the Higgs boson can be expected without
additional input.

New results about $CP$ violation are obtained: ``indirect'' $CP$ violation
just appears to be the consequence of $P$ violation (unitarity compels the
electroweak mass eigenstates to be $C$ eigenstates), and the role of the
Kobayashi-Maskawa mixing matrix for fermions fades away: mesonic mass
eigenstates can still be $CP$ eigenstates despite the presence of a complex
phase.

I briefly show how the customary results for the leptonic decays of
pseudoscalar mesons are recovered.

The extension to leptons is discussed. It is shown, in the context of
previous works linking the effective $V-A$ structure of weak currents to the
non-observation of right-handed neutrinos and the masslessness of the
left-handed ones, how the custodial symmetry
constrains them to be Majorana particles.
}
\vfill
\null\hfil\epsffile{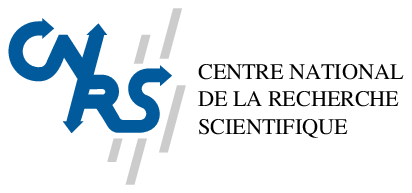}
\end{titlepage}
\section{Introduction.}
\label{section:introduction}

Many difficulties in hadronic physics stem from the fields in the
Lagrangian (quarks) not being the observed particles or asymptotic states
(mesons, baryons). 

I show here that the standard model of electroweak interactions
\cite{GlashowSalamWeinberg} can
be straightforwardly extended to $J=0$ mesons \cite{Machet1},
and I will put a special emphasis on its symmetry properties.

They concern:\l
- the occurrence of a chiral $SU(2)_L \times SU(2)_R$ chiral symmetry,
explicitly broken, when the hypercharge coupling $g'$ is non vanishing,
and spontaneously broken, when the Higgs boson gets a non-vanishing vacuum
expectation value, to a ``custodial'' diagonal $SU(2)_V$ symmetry
\cite{Machet1}; the
latter is  local when $g' =0$. It includes the electric charge 
generator, such that the quantization of the electric charge is directly
linked with the custodial symmetry staying unbroken.
It is put in relation with the ideas of electric-magnetic duality
\cite{Olive}, and the occurrence of a strongly interacting
sector in the standard model;\l
- the transformation properties by $CP$: I show \cite{Machet2}
that ``indirect'' $CP$ violation is only a consequence of $P$ violation, and
that the existence of a complex phase in the mixing matrix  is no longer a
sufficient condition for electroweak mass eigenstates to be different from
$CP$ eigenstates.

The electroweak group $SU(2)_L \times U(1)$ is embedded into $U(N)_L \times
U(N)_R$ in such a way that it acts on fermions like in the 
Glashow-Salam-Weinberg model; the embedding is accordingly characterized by the
Cabibbo-Kobayashi-Maskawa \cite{CabibboKobayashiMaskawa} unitary matrix.

This extension of the standard model provides new ideas concerning chiral 
symmetry in the physics of mesons and  also has consequences on the way
we interpret their spectrum.

The spontaneous breaking, by $\la H \ra \not = 0$, of the chiral 
$SU(2)_L \times SU(2)_R$ down to $SU(2)_V$ and of $SU(2)_L \times U(1)$ 
down to the electromagnetic $U(1)_{em}$ are the same phenomenon. The three
corresponding Goldstone bosons, that become the longitudinal components
of the three massive gauge
fields, are linear combinations of the known pseudoscalar mesons (pions,
kaons $\ldots$); the latter (pions, kaons $\ldots$)
 are naturally massive since they only coincide
with the Goldstones of the broken chiral symmetry in the case of one
generation and, of course, vanishing mixing angle. 
In the real case of three generations, no $J=0$ meson which is a flavour
eigenstate is to be interpreted as a Goldstone particle.

The chiral breaking relevant to the physics of $J=0$ mesons thus
appears to be that
of $SU(2)_L \times SU(2)_R$ into $SU(2)_V$ rather than the one of 
$U(N)_L \times U(N)_R$ into the diagonal $U(N)$ flavour subgroup
\cite{CurrentAlgebra} ($N$ is the number of flavours).
$SU(2)_V$ allows  $N^2$ independent mass
scales, twice the number  shown to be allowed by the chiral
$SU(2)_L \times SU(2)_R$ and the electroweak $SU(2)_L \times
U(1)$ symmetries. The doubling from $N^2/2$ to $N^2$ can in particular 
split scalar and pseudoscalar mesons, though the electroweak mass eigenstates 
do not have in general a definite parity.

The spectrum of $J=0$ mesons has acquired the same arbitrariness as the one 
of the leptons. 
It includes the Higgs boson, the mass of which is now not
more but also not less explained than the ones of the other mesons.

A reduction of the number of arbitrary mass parameters would need an
additional symmetry.

The quantization of the electric charge for all asymptotic states requires
that the same formalism be applied to leptons \cite{Machet3}.
Since the hadronic sector is
now naturally anomaly-free (no fermionic field is involved any longer),
we cannot invoke any longer the usual cancelation
\cite{BouchiatIliopoulosMeyer} between quarks and leptons.
The natural candidate is the purely vectorial theory described in
\cite{BellonMachet}, in which the introduction of a composite triplet of
scalars links, at leading order in $1/N$, the ``decoupling'', by an exact
``see-saw'' mechanism, of an
infinitely massive right-handed neutrino, to the effective $V-A$ structure of
weak currents. The observed neutrino is then exactly massless.
It is shown in \cite{Machet3} that the existence of the same
custodial $SU(2)_V$ symmetry compels the neutrino to be a Majorana
particle.

\section{The chiral group $\mathbf{U(N)_L \times U(N)_R}$ and the electroweak 
subgroup $\mathbf{SU(2)_L \times U(1)}$.}
\label{section:UN}

Let $N/2$ be the number of generations; the number $N$ of
``flavours'' is even (the construction below cannot be performed for $N$
odd). The observed $J=0$ mesons are generally classified according to
their parity quantum number $P= \pm 1$ and it is convenient to introduce
the parity changing operator $\cal P$ which transforms a scalar into a
pseudoscalar and {\em vice-versa}; a $\cal P$-even meson will be later
written (see subsection \ref{subsec:reps})
as the sum ``scalar +  pseudoscalar'', and a $\cal P$-odd meson as
the  difference ``scalar - pseudoscalar''. The laws of transformation of
mesons by the chiral group $U(N)_L \times U(N)_R$ are indeed most simply
expressed in terms of $\cal P$-even and $\cal P$-odd particles.

Both types are taken to be $N\times N$ matrices $\mathbb M$, which will be 
given an index ``{\em odd}'' or ``{\em even}''. 
The quarks are here only considered as mathematical entities
\cite{GellMann}, and the mesons
fields as objects transforming as composite quark-anti quark operators.
The link with physically observed particles is more thoroughly examined in
section \ref{section:mesons}.
The link with composite
quark-anti quark operators is made by sandwiching $\mathbb M$ between
the $N$-vector of quarks $\Psi$ in the fundamental representation of $U(N)$
\begin{equation}
\Psi =
\bpma  u\\ c\\ \vdots\\  d\\ s\\ \vdots \epma
.
\label{eq:Psi}
\end{equation}
and its hermitian conjugate, and introducing, according to the
transformation by parity, the appropriate $\gamma_5$ matrix.

\subsection{The action of $\mathbf{U(N)_L \times U(N)_R}$.}
\label{subsec:action}

A generator $\Bbb A$ of $U(N)_L \times U(N)_R$ is a set of two $N\times N$
matrices $({\Bbb A}_L, {\Bbb A}_R)$. A generator of a diagonal subgroup
satisfies ${\Bbb A}_L =  {\Bbb A}_R$.

At the level of the algebra, we define the action of the {\em left} and {\em
right} generators by:

\vbox{
\bea
{\Bbb A}^i_L\,.\, {\Bbb M}_{even} &\stackrel{def}{=}&
            - \,{\Bbb A}^i_L \,{\Bbb M}_{even}
                                           ={1\over 2} \left(
            [{\Bbb M}_{{\cal P}even},{\Bbb A}^i_L]
                            - \{{\Bbb M}_{even},{\Bbb A}^i_L\}
                                    \right),\cr
{\Bbb A}^i_L\,.\, {\Bbb M}_{odd} &\stackrel{def}{=}&
            +\,{\Bbb M}_{odd}\,{\Bbb A}^i_L
                                           =\hskip 5mm {1\over 2} \left(
            [{\Bbb M}_{odd},{\Bbb A}^i_L]
                            + \{{\Bbb M}_{odd},{\Bbb A}^i_L\}
                                    \right),\cr
{\Bbb A}^i_R\,.\, {\Bbb M}_{even} &\stackrel{def}{=}&
            +\,{\Bbb M}_{even} \,{\Bbb A}^i_R
                                           ={1\over 2} \left(
            [{\Bbb M}_{even},{\Bbb A}^i_R]
                            + \{{\Bbb M}_{even},{\Bbb A}^i_R\}
                                    \right),\cr
{\Bbb A}^i_R\,.\, {\Bbb M}_{odd} &\stackrel{def}{=}&
            - \,{\Bbb A}^i_R \,{\Bbb M}_{odd}
                                          =\hskip 5mm{1\over 2} \left(
            [{\Bbb M}_{odd},{\Bbb A}^i_R]
                            - \{{\Bbb M}_{odd},{\Bbb A}^i_R\}
                                    \right),
\label{eq:actionUN}
\eea
}
which is akin to left- and right- multiplying $N\times N$ matrices.

At the level of the group, let ${\cal U}_L \times {\cal U}_R$ be a finite
transformation of the chiral group; we have
\bea
{\cal U}_L \times {\cal U}_R \,.\, {\Bbb M}_{even}  &=&
            {\cal U}_L^{-1}  \,{\Bbb M}_{even} \,{\cal U}_R, \cr
{\cal U}_L \times {\cal U}_R \,.\, {\Bbb M}_{odd}  &=&
            {\cal U}_R^{-1} \,{\Bbb M}_{odd} \,{\cal U}_L,
\label{eq:UNg}
\eea
reminiscent of the group action in a $\sigma$-model \cite{GellMannLevy}
\cite{CurrentAlgebra}
with a $U(N)_L \times U(N)_R$ group of symmetry.
Note that  ``left'' and ``right'' are swapped in the
action on the $\cal P$-odd scalars with respect to the $\cal P$-even ones.

The actions defined above can be derived straightforwardly by acting with
the left and right $U(N)$ groups on the fermionic ``components'' of 
$\bar\Psi (1\pm\gamma_5){\Bbb M}\Psi$: the left-handed generators are 
then given a $(1-\gamma_5)/2$ projector, and the right ones a $(1+\gamma_5)/2$. 
That the $\gamma_5$ of the projectors has to go through the $\gamma^0$ of
$\bar \Psi$ yields both commutators and anticommutators in
eq.~(\ref{eq:actionUN}).

\subsection{The electroweak $\mathbf{SU(2)_L \times U(1)}$.}
\label{subsec:SU2}

The extension of the Glashow-Salam-Weinberg model \cite{GlashowSalamWeinberg}
to $J=0$ mesons proposed in \cite{Machet1} is
a $SU(2)_L \times U(1)$ gauge theory of matrices. As the action of
the gauge group can only be defined if its generators are also $N \times N$
matrices, it is considered as a subgroup of the chiral group. Its
orientation within the latter has to be compatible with the customary action
of the electroweak group on fermions, and is determined by a unitary $N/2
\times N/2$ matrix which is nothing else than the Cabibbo-Kobayashi-Maskawa
mixing matrix $\Bbb K$ \cite{CabibboKobayashiMaskawa}.

We hereafter decompose all $N \times N$ matrices into $N/2 \times N/2$
blocks.

The $SU(2)_L$ generators are
\begin{equation}
{\Bbb T}^3_L = {1\over 2}\left(\begin{array}{rrr}
                        {\Bbb I} & \vline & 0\\
                        \hline
                        0 & \vline & -{\Bbb I}
\end{array}\right),\
{\Bbb T}^+_L =           \left(\begin{array}{ccc}
                        0 & \vline & {\Bbb K}\\
                        \hline
                        0 & \vline & 0           \end{array}\right),\
{\Bbb T}^-_L =           \left(\begin{array}{ccc}
                        0 & \vline & 0\\
                        \hline
                        {\Bbb K}^\dagger & \vline & 0
\end{array}\right);
\label{eq:SU2L}
\end{equation}
they act trivially on the $N$-vector of quarks $\Psi$ (they are then given a
left $(1-\gamma_5)/2$ projector) in the same way as in the
Glashow-Salam-Weinberg model, ensuring the consistency of our approach with
the latter.
 
${\Bbb T}^+$ and ${\Bbb T}^-$ stand respectively for $({\Bbb T}^1 +
i~{\Bbb T}^2)$ and $({\Bbb T}^1 - i~{\Bbb T}^2)$.
$\Bbb I$ is the $N/2 \times N/2$ identity matrix.

The hypercharge $U(1)$ generator satisfies the Gell-Mann-Nishijima relation 
\cite{GellMannNishijima} (written in its ``chiral'' form)
\begin{equation}
({\Bbb Y}_L,{\Bbb Y}_R) = ({\Bbb Q}_L,{\Bbb Q}_R) - ({\Bbb T}^3_L,0);
\label{eq:GMN}\end{equation}
it is non-diagonal and commutes with $SU(2)_L$.
Taking the customary expression for the electric charge operator
\begin{equation}
{\Bbb Q} = \left(\begin{array}{ccc}
                        2/3 & \vline & 0\cr
                        \hline
                        0 & \vline & -1/3
           \end{array}\right),
\label{eq:Q}
\end{equation}
yields back the usual formula for the ``left'' and ``right''
hypercharges
\begin{equation}
{\Bbb Y}_L = {1\over 6}{\Bbb I}, \quad {\Bbb Y}_R = {\Bbb Q}_R.
\end{equation}
The ``alignment'' of the electroweak subgroup inside the chiral group is
controlled by a unitary matrix, $({\Bbb R},{\Bbb R})$, acting diagonally,
with
\begin{equation}
{\Bbb R} =             \left(\begin{array}{ccc}
                        {\Bbb I} & \vline & 0\\
                        \hline
                        0 & \vline & {\Bbb K}           \end{array}\right),
\label{eq:rotation}
\end{equation}
The electroweak group defined by eq.~(\ref{eq:SU2L}) is the one with generators
\begin{equation}
{\Bbb R}^\dagger \vec t_L\; {\Bbb R}\; ;
\label{eq:rotgroup}
\end{equation}
with 
\begin{equation}
{t}^3_L = {1\over 2}\left(\begin{array}{ccc}
                        {\Bbb I} & \vline & 0\\
                        \hline
                        0 & \vline & -{\Bbb I}
\end{array}\right),\
{t}^+_L =           \left(\begin{array}{ccc}
                        0 & \vline & {\Bbb I}\\
                        \hline
                        0 & \vline & 0           \end{array}\right),\
{t}^-_L =           \left(\begin{array}{ccc}
                        0 & \vline & 0\\
                        \hline
                        {\Bbb I} & \vline & 0           \end{array}\right).
\label{eq:generic}
\end{equation}
In practice, this rotation only acts on the ${t}^\pm$ generators
(we require ${t}^- = ({t}^+)^\dagger$, such that the
unit matrices in eqs.~(\ref{eq:generic},\ref{eq:SU2L}) have the same dimension).

\subsection{The electroweak representations of $\mathbf{J=0}$ mesons.}
\label{subsec:reps}

In the same way (see eq.~(\ref{eq:actionUN})) as we wrote the action of the
chiral group on scalar fields represented by $N \times N$ matrices $\Bbb M$,
we define the action of its $SU(2)_L$ subgroup, to which we add the action
of the electric charge $\Bbb Q$ according to:
\begin{equation}
{\Bbb Q}\,.\,{\Bbb M} =  [{\Bbb M},{\Bbb Q}];
\label{eq:charge1}
\end{equation}
it acts by commutation because it is a diagonal operator (see subsection
\ref{subsec:action}).

The representations of the electroweak group $SU(2)_L \times U(1)$
are  also of two types, $\cal P$-even and $\cal P$-odd,
according to their transformation properties by the parity changing operator
$\cal P$. Only representations transforming alike can be
linearly mixed to form another representation of the same type.

We can build a very special type of representations,
in the form of quadruplets $({\Bbb M}^0, \vec {\Bbb M})$, 
where the $\Bbb M$'s are still $N \times N$ matrices;
$\vec {\Bbb M}$ stands for the sets of complex matrices
 $\{{\Bbb M}^1, {\Bbb M}^2, {\Bbb M}^3\}$ or
$\{{\Bbb M}^3, {\Bbb M}^+, {\Bbb M}^-\}$ with ${\Bbb M}^+ = ({\Bbb M}^1 +
i\,
{\Bbb M}^2)/\sqrt{2}\;,\; {\Bbb M}^- = ({\Bbb M}^1 - i\, {\Bbb
M}^2)/\sqrt{2}$.

Let us consider quadruplets of the form

\hskip -.5cm\vbox{
\bea
& &\Phi(\Bbb D)=
({\Bbb M}\,^0, {\Bbb M}^3, {\Bbb M}^+, {\Bbb M}^-) (\Bbb D)\cr =
& & \left[
 {1\over \sqrt{2}}\left(\begin{array}{ccc}
                        {\Bbb D} & \vline & 0\\
                        \hline
                      0 & \vline & {\Bbb K}^\dagger\,{\Bbb D}\,{\Bbb K}
                   \end{array}\right),
{i\over \sqrt{2}} \left(\begin{array}{ccc}
                        {\Bbb D} & \vline & 0\\
                        \hline
                    0 & \vline & -{\Bbb K}^\dagger\,{\Bbb D}\,{\Bbb K}
                   \end{array}\right),
i\left(\begin{array}{ccc}
                        0 & \vline & {\Bbb D}\,{\Bbb K}\\
                        \hline
                        0 & \vline & 0           \end{array}\right),
 i\left(\begin{array}{ccc}
                        0 & \vline & 0\\
                        \hline
                        {\Bbb K}^\dagger\,{\Bbb D} & \vline & 0
                    \end{array}\right)
             \right].\cr
& &
\label{eq:quad}
\eea}

where $\Bbb D$ is a real $N/2 \times N/2$ matrix.

That the entries ${\Bbb M}^+$ and ${\Bbb M}^-$ are, up to a sign,
hermitian conjugate ({\em i.e.} charge conjugate) requires that the $\Bbb
D$'s are restricted to symmetric or antisymmetric matrices.

Because of the presence of an ``$i$'' for the for ${\Bbb M}^{3,\pm}$ and not
for ${\Bbb M}^0$, the quadruplets always mix entries of different behaviour
by hermitian (charge) conjugation, and are consequently not hermitian
representations.

The action of $SU(2)_L \times U(1)$ on these quadruplets is defined by its
action on each of the four components, as written  in
eqs.~(\ref{eq:actionUN},\ref{eq:charge1}).
It turns out that it can be rewritten in the form
(the Latin indices $i,j,k$ run from $1$ to $3$):
\bea
{\Bbb T}^i_L\,.\,{\Bbb M}^j_{even} &=& -{i\over 2}\left(
              \epsilon_{ijk} {\Bbb M}^k_{even} +
                           \delta_{ij} {\Bbb M}_{even}^0
                              \right),\cr
{\Bbb T}^i_L\,.\,{\Bbb M}_{even}^0 &=&
                                {i\over 2}\; {\Bbb M}_{even}^i;
\label{eq:actioneven}
\eea
and
\bea
{\Bbb T}^i_L\,.\,{\Bbb M}_{odd}^j &=& -{i\over 2}\left(
                   \epsilon_{ijk} {\Bbb M}_{odd}^k -
                           \delta_{ij} {\Bbb M}_{odd}^0
                              \right),\cr
{\Bbb T}^i_L\,.\,{\Bbb M}_{odd}^0 &=&
                        \hskip 5mm  -{i\over 2}\; {\Bbb M}_{odd}^i.
\label{eq:actionodd}
\eea
The charge operator acts indifferently on ${\cal P}$-even and ${\cal P}$-odd
matrices by:
\bea
{\Bbb Q}\,.\,{\Bbb M}\,^i &=& -i\,\epsilon_{ij3} {\Bbb M}\,^j,\cr
{\Bbb Q}\,.\,{\Bbb M}\,^0 &=& 0\,,
\label{eq:charge2}
\eea
and the action of the $U(1)$ generator $\Bbb Y$ follows from
eq.~(\ref{eq:GMN}).

Still as a consequence of (\ref{eq:actionUN}), the action of the
``right'' group $SU(2)_R$ is of the same form as displayed in
eqs.~(\ref{eq:actioneven},\ref{eq:actionodd}) but with the signs in front of
the ${\Bbb M}^0$'s all swapped.

Taking the hermitian conjugate of any representation $\Phi$ swaps the
relative sign between ${\Bbb M}^0$ and $\vec{\Bbb M}$; as a consequence,
$\Phi^\dagger_{even}$ transforms by $SU(2)_L$
as would formally do a ${\cal P}$-odd representation, and {\em vice-versa};
on the other hand, the quadruplets (\ref{eq:quad}) are also representations
of $SU(2)_R$, the action of which is obtained by swapping
eqs.~(\ref{eq:actioneven}) and (\ref{eq:actionodd});
so, the hermitian conjugate of a given representation of $SU(2)_L$ is a
representation of $SU(2)_R$ with the same law of transformation, and
{\em vice-versa}. The same result holds for any (complex) linear
representation $U$ of quadruplets transforming alike by the gauge group.

We see that we now deal with 4-dimensional representations of $SU(2)_L
\times U(1)$, which are also, by the above remark, representations of
$SU(2)_R$.
In the basis formed by the four entries of any such representation,
the generators of the electroweak
group can be rewritten as $4 \times 4$ matrices. This is also the case for
the generators of the diagonal  $SU(2)_V$.

They decompose into ``symmetric'' representations, corresponding to
$\; \Bbb D = \,{\Bbb D}^\dagger$, and ``antisymmetric'' ones for which
$\; \Bbb D = -\,{\Bbb D}^\dagger$.

There are $N/2(N/2 +1)/2$ independent real symmetric $\Bbb D$ matrices;
hence, the sets of ``even'' and ``odd'' symmetric quadruplet representations
of the type (\ref{eq:quad}) both have dimension $N/2(N/2 +1)/2$.
Similarly, the antisymmetric ones form two sets of dimension $N/2(N/2
-1)/2$.

Every representation above is a reducible representation of $SU(2)_L$ (or
$SU(2)_R$)
and is the sum of two (complex) representations of spin $1/2$.
This makes it isomorphic to the standard scalar set of the
Glashow-Salam-Weinberg model \cite{GlashowSalamWeinberg}.

Now, if we consider the transformation properties by the diagonal $SU(2)_V$,
all $\vec {\Bbb M}$'s are (spin $1$) triplets, lying in the adjoint
representation, while all ${\Bbb M}^0$'s are singlets.

By adding or subtracting eqs.~(\ref{eq:actioneven}) and
(\ref{eq:actionodd}),
and defining scalar ($\Bbb S$) and pseudoscalar ($\Bbb P$) fields by
\begin{equation}
({\Bbb M}_{even} + {\Bbb M}_{odd}) = {\Bbb S},
\label{eq:scalar}
\end{equation}
and
\begin{equation}
({\Bbb M}_{even} - {\Bbb M}_{odd}) = {\Bbb P},
\label{eq:pseudo}
\end{equation}
one finds two new types of stable quadruplets which include objects of
different
parities, but which now correspond to a given $CP$ quantum number, depending
in particular whether $\Bbb D$ is a symmetric or skew-symmetric matrix
\begin{equation}
({\Bbb M}\,^0, \vec {\Bbb M}) = ({\Bbb S}^0, \vec {\Bbb P}),
\label{eq:SP}
\end{equation}
and
\begin{equation}
({\Bbb M}\,^0, \vec {\Bbb M}) = ({\Bbb P}\,^0, \vec {\Bbb S});
\label{eq:PS}
\end{equation}
they both transform by the gauge group like $\cal P$-even reps, according to
eq.~(\ref{eq:actioneven}), and thus can be linearly mixed.  As they span the
whole space of $J=0$ mesons too, this last property makes them
specially convenient to consider.

By hermitian conjugation, that is charge conjugation,
 a ``symmetric'' $({\Bbb M}\,^0, \vec {\Bbb M})$
representation gives $({\Bbb M}\,^0, -\vec {\Bbb M})$; an ``antisymmetric''
representation gives $(-{\Bbb M}\,^0, \vec {\Bbb M})$; the representations
(\ref{eq:SP}) and (\ref{eq:PS}) are consequently
representations of given $CP$\ (charge\ conjugation\ $\times$\  parity):
``symmetric'' $({\Bbb S}^0, \vec {\Bbb P})$'s and ``antisymmetric''
$({\Bbb P}\,^0, \vec {\Bbb S})$'s are $CP$-even, while ``symmetric''
$({\Bbb P}\,^0, \vec {\Bbb S})$'s and ``antisymmetric''
$({\Bbb S}^0, \vec {\Bbb P})$'s are $CP$-odd.

\subsection{``Strong'' and electroweak basis for the mesons.}
\label{subsec:basis}

We call ``strong'' basis the set of flavour (and parity) $U(N)$ eigenstates.
They are represented by $N^2$ matrices ${\Bbb F}^{ij}$ for scalars, and
${\Bbb F}^{ij}_5, \ i,j=1\cdots N$ for pseudoscalars, in which 
only one entry, the one at the crossing of the $i$th line and the $j$th column,
is non vanishing and has the value $1$. This is equivalent, in the quark
language, to the set of $\bar q_i  q_j$ and $\bar q_i \gamma_5 q_j$ states.
The most general meson $\Bbb M$  thus decomposes on the strong basis 
according to
\begin{equation}
{\Bbb M} = \sum_{i,j=1\cdots N} M_{ij} {\Bbb F}^{ij} + M_{ij}^5 {\Bbb
F}^{ij}_5.
\label{eq:decomp}
\end{equation}
A quadratic expression we call diagonal in the basis of strong eigenstates
if it only involves tensor products of the type ${\Bbb F}^{ij} \otimes
{\Bbb F}^{ji}$ and ${\Bbb F}^{ij}_5 \otimes {\Bbb F}^{ji}_5$
(we use hereafter the notation $\otimes$ for the tensor product of two 
fields, not to be mistaken with the ordinary product of matrices).

\section{The $\mathbf{SU(2)_L \times U(1)}$ Lagrangian.}
\label{section:lagrangian}

Having defined the fundamental fields and how they transform by
the groups of symmetries involved in the problem, we shall now explicitly
write the $SU(2)_L \times U(1)$ gauge Lagrangian for $J=0$ mesons.
It requires knowing which polynomial expressions
are invariant by the gauge group (in practice we need only quadratic
invariants; the quartic invariants are constructed as products of any two
quadratic ones and higher powers are forbidden by the requirement of
renormalizability).

It is from the nature of these invariants that the chiral structure 
of our construction and the role of the group
$SU(2)_L \times SU(2)_R$, which will be examined in detail in the next
section, spring out.

\subsection{The quadratic invariants.}
\label{subsec:invariants}

To every representation is associated a unique quadratic expression invariant
by the electroweak gauge group $SU(2)_L \times U(1)$
\begin{equation}
{\cal I} = ({\Bbb M}^0, \vec {\Bbb M})\otimes ({\Bbb M}^0, \vec {\Bbb M})=
 {\Bbb {\Bbb M}}\,^0 \otimes {\Bbb {\Bbb M}}\,^0 +
                 \vec {\Bbb M} \otimes \vec {\Bbb M};
\label{eq:invar}
\end{equation}
$\vec {\Bbb M} \otimes \vec {\Bbb M}$
stands for $\sum_{i=1,2,3} {\Bbb M}\,^i \otimes  {\Bbb M}\,^i$.

Other invariants can be built like tensor products of two representations
transforming alike by the gauge group: two $\cal P$-odd or two $\cal P$-even,
two $({\Bbb S}^0,\vec {\Bbb P})$, two $({\Bbb P}^0,\vec {\Bbb S})$, or one
$({\Bbb S}^0,\vec {\Bbb P})$ and one $({\Bbb P}^0,\vec {\Bbb S})$; for
example such is
\begin{equation}
{\cal I}_{1\ti 2} = ({\Bbb S}^0,\vec {\Bbb P})({\Bbb D}_1) \otimes
                ({\Bbb P}^0,\vec {\Bbb S})({\Bbb D}_2)
 ={\Bbb S}^0({\Bbb D}_1) \otimes {\Bbb P}^0({\Bbb D}_2) +
                  \vec {\Bbb P}({\Bbb D}_1) \otimes \vec {\Bbb S}({\Bbb
D}_2).
\label{eq:I12}\end{equation}
According to the remark made in the previous section, all the above
expressions are also invariant by the action of $SU(2)_R$.

\subsection{A special combination of invariants.}
\label{subsec:special}

For the relevant cases $N=2,4,6$, there exists a set of $\Bbb D$ matrices
such that the algebraic sum (specified below)
of invariants extended over all  representations defined by
(\ref{eq:SP},\ref{eq:PS},\ref{eq:quad})

\hskip -1cm\vbox{
\bea
&&
{1\over 2}
\left((\sum_{symmetric\ {\Bbb D}} - \sum_{skew-symmetric\ {\Bbb D}})
\left( ({\Bbb S}^0, \vec {\Bbb P})({\Bbb D})
                     \otimes  ({\Bbb S}^0, \vec {\Bbb P})({\Bbb D})
- ({\Bbb P}^0, \vec {\Bbb S})({\Bbb D})
                     \otimes  ({\Bbb P}^0, \vec {\Bbb S})({\Bbb D})
\right)\right)\cr
&=&
{1\over 4}
\left((\sum_{symmetric\ {\Bbb D}} - \sum_{skew-symmetric\ {\Bbb D}})
\left(\Phi_{even}({\Bbb D})\otimes\Phi^\dagger_{odd}({\Bbb
D})
  +\Phi_{odd}({\Bbb D})\otimes \Phi^\dagger_{even}({\Bbb D})
\right) \right)\cr
&&
\label{eq:Idiag}\eea
}

is diagonal both in the electroweak basis and in the basis of
strong eigenstates:
in the latter basis, all terms are normalized alike to $(+1)$
(including the sign).
Note that two ``$-$'' signs  occur in eq.~(\ref{eq:Idiag})
\footnote{Eq.~(\ref{eq:Idiag} specifies eq.~(25) of
\cite{Machet1}, in which the ``$-$'' signs were not explicitly written.}
:\l
- the first between the $({\Bbb P}^0, \vec{\Bbb S})$ and
$({\Bbb S}^0, \vec{\Bbb P})$ quadruplets, because, as seen on
eq.~(\ref{eq:quad}), the ${\Bbb P}^0$ entry of the former has no ``$i$''
factor, while the $\vec{\Bbb P}$'s of the latter do have one; as we define
all pseudoscalars without an ``$i$''
(like $\pi^+ = \bar u d$), a $(\pm i)$ relative factor has to be introduced
between the two types of representations, yielding a ``$-$'' sign in
eq.~(\ref{eq:Idiag});\l
- the second for  the representations corresponding to skew-symmetric
$\Bbb D$ matrices, which have an opposite
behaviour by charge conjugation ({\em i.e.} hermitian conjugation)
as compared to the ones with symmetric ${\Bbb D}$'s.

The $SU(2)_L \times U(1)$ kinetic Lagrangian for $J=0$ mesons
is built from the special combination  of invariants
(\ref{eq:Idiag}), now used for the covariant
derivatives of the fields with respect to the gauge group;
its part involving pure derivatives is thus diagonal in both the strong
and electroweak basis, too.

The characteristic property of the combination (\ref{eq:Idiag})
is most simply verified for the ``non-rotated''
$SU(2)_L \times U(1)$ group and representations \cite{Machet1}.
Explicitly \cite{Machet2}:

\subsubsection{$N=2.$}

It is a trivial case: $\Bbb D$ is a number.

\subsubsection{$N=4.$}

The four $2\times 2$ $\Bbb D$ matrices ($3$ symmetric and $1$
skew-symmetric) can be taken as
\begin{equation}
{\Bbb D}_1 = \left( \ba{cc} 1 & 0 \cr
                            0 & 1     \ea \right),\
{\Bbb D}_2 = \left( \ba{rr} 1 & 0 \cr
                            0 & -1    \ea \right),\
{\Bbb D}_3 = \left( \ba{cc} 0 & 1 \cr
                            1 & 0     \ea \right),\
{\Bbb D}_4 = \left( \ba{rr} 0 & 1 \cr
                           -1 & 0     \ea \right).
\label{eq:N=4}
\end{equation}

\subsubsection{$N=6.$}\label{subsubsec:N=6}

The nine $3 \times 3$ $\Bbb D$ matrices ($6$ symmetric and $3$ skew-symmetric),
can be taken as

\hskip -1cm\vbox{
\bea
& &{\Bbb D}_1 = \sqrt{{2\over 3}}\left( \ba{ccc}
                                1  &  0  &  0  \cr
                                0  &  1  &  0  \cr
                                0  &  0  &  1 \ea \right), \cr
& &{\Bbb D}_2 ={2\over\sqrt{3}} \left( \ba{ccc}
                \sin\alpha  &     0    &    0    \cr
       0     & \sin(\alpha\pm{2\pi\over 3})&   0   \cr
       0     &                    0        & \sin(\alpha\mp{2\pi\over 3})
       \ea\right),\
{\Bbb D}_3 ={2\over\sqrt{3}} \left( \ba{ccc}
                \cos\alpha  &     0    &    0    \cr
       0     & \cos(\alpha\pm{2\pi\over 3})&   0   \cr
       0     &                    0        & \cos(\alpha\mp{2\pi\over 3})
       \ea\right), \cr
& & {\Bbb D}_4 =\left( \ba{ccc}
                                0  &  0  &  1 \cr
                                0  &  0  &  0 \cr
                                1  &  0  &  0   \ea \right),\
     {\Bbb D}_5 =\left( \ba{rrr}
                                0  &  0  &  1 \cr
                                0  &  0  &  0 \cr
                               -1  &  0  &  0   \ea \right),\cr
& & {\Bbb D}_6 = \left( \ba{ccc}
                                0  &  1  &  0  \cr
                                1  &  0  &  0  \cr
                                0  &  0  &  0   \ea \right), \
 {\Bbb D}_7 = \left( \ba{rrr}
                                0  &  1  &  0  \cr
                               -1  &  0  &  0  \cr
                                0  &  0  &  0   \ea \right), \
 {\Bbb D}_8 = \left( \ba{ccc}
                                0  &  0  &  0  \cr
                                0  &  0  &  1  \cr
                                0  &  1  &  0   \ea \right), \
 {\Bbb D}_9 = \left( \ba{rrr}
                                0  &  0  &  0  \cr
                                0  &  0  &  1  \cr
                                0  & -1  &  0   \ea \right),\cr
& &
\label{eq:N=6}
\eea
}

where $\alpha$ is an arbitrary  phase.

{\em Remark}: as ${\Bbb D}_1$ is the only matrix with a non vanishing trace,
${\Bbb S}^0({\Bbb D}_1)$ is the only neutral scalar matrix with the
same property;  we take it as the Higgs boson.

Considering that it is the only scalar with a non-vanishing vacuum
expectation value prevents the occurrence of a hierarchy problem
\cite{GildenerWeinberg}.

This last property is tantamount, in the ``quark language'', to taking
the same value for all condensates $\la\bar q_i q_i\ra, i=1 \cdots N$,
in agreement with the flavour independence of ``strong interactions''
between fermions, supposedly at the origin of this phenomenon in the
traditional framework.

As the spectrum of mesons is, in the present model, disconnected from a
hierarchy between quark condensates (see below),
it is not affected by our choice of a single Higgs boson.
 
\subsection{The basic property of the quadratic invariants.}
\label{subsec:property}

The quadratic $SU(2)_L$ invariants are not {\em a priori} self conjugate
expressions 
and have consequently  no definite property by hermitian conjugation;
in particular, the one associated with a most general representation $U$ is
$U \otimes U$ and {\em not} $U \otimes U^\dagger$ (we have seen in the
previous section that $U$ and $U^\dagger$ do not transform alike
by the gauge group).

As far as one only deals with representations of the type of
eqs.~(\ref{eq:SP},\ref{eq:PS}), like in the special
combination (\ref{eq:Idiag}), it has no consequence since
each of their entries has a well defined behaviour by hermitian conjugation
and the associated quadratic invariants are then always hermitian.

But electroweak mass eigenstates are in general (complex)
linear combinations of reps (\ref{eq:SP},\ref{eq:PS}) and have,
consequently, no definite behaviour by hermitian (charge) conjugation.
This has consequences, in particular as far as the transformation
properties by $CP$ are concerned (see section \ref{section:CP}).

\subsection{First remarks on the spectrum of $\mathbf{J=0}$ mesons.} 
\label{subsec:remark}

The quadratic invariants are used to build the gauge invariant mass terms 
in the Lagrangian.

As long as $SU(2)_L \times U(1)$ is unbroken, there are {\em a priori} 
as many $(N^2/2)$ independent mass scales as there are independent 
representations. 
They share with the leptonic case the same arbitrariness.

Since there are eleven pseudoscalar mesons which ``include'' the quark top,
and since they cannot all be fitted in a unique representation, it should not
be a surprise if different mass scales are found to correspond to ``topped''
mesons; one should be aware not to misinterpret them,
like by advocating for the occurrence of a new generation.

The number of mass scales could be reduced if the theory has additional
symmetries.

Note that, from the diagonalization property of eq.~(\ref{eq:Idiag}),
identical mass terms for the $({\Bbb S}^0, \vec{\Bbb P}) (\Bbb D), 
({\Bbb P}^0, \vec{\Bbb S}) (\Bbb D)$ mutliplets correspond to the same property
for the flavour eigenstates.

Whatever convenient be the distinction between scalars and pseudoscalars,
one must keep in mind that, in a parity violating theory such as ours, the most
general electroweak mass eigenstates do not have a definite parity.

\section{The chiral symmetry and its breaking to the custodial
$\mathbf{SU(2)_V}$.}
\label{section:chiral}

This section is devoted to the study of the symmetries of the
$SU(2)_L \times U(1)$ Lagrangian for $J=0$ mesons.
The origin of the ``custodial'' $SU(2)_V$ symmetry as the result of the
breaking of a chiral $SU(2)_L \times SU(2)_R$ symmetry is made explicit,
together with the similarity of the chiral and electroweak breaking.

Further consequences on the spectrum of mesons and the nature of the
Goldstones bosons are emphasized.

Finally, the tight link between the custodial symmetry and the quantization
of the electric charge is examined in the light of electric-magnetic
duality and the recent works by Cho, Maison and Kimm
\cite{ChoMaison}\cite{ChoKimm}.

\subsection{The chiral $\mathbf{SU(2)_L \times SU(2)_R}$ symmetry.}
\label{subsec:chiralSU2}

All $SU(2)_L \times U(1)$ quadratic invariants that are used to build the
Lagrangian  are also invariant by $SU(2)_R$. The scalar potential is thus
$SU(2)_L \times SU(2)_R$ chirally invariant.

Because the coupling constant $g'$ of the hypercharge $U(1)$ is different
from the $SU(2)_L$ coupling $g$, only the covariant derivatives of the
fields with respect to $SU(2)_L$ have definite transformation properties
with respect to $SU(2)_R$, which are the same as the fields themselves,
{\em when the right-handed $\vec W$'s are identified with the left-handed
ones}. This can be done since the laws of transformations for the adjoint
representations of both groups are identical.

The weak hypercharge group breaks this symmetry, as expected by the
Gell-Mann-Nishijima relation which shows that it is ``polarized''.

We thus conclude that {\em the $SU(2)_L \times U(1)$ Lagrangian for $J=0$
mesons has a chiral $SU(2)_L \times SU(2)_R$ symmetry at the limit $g'
\rightarrow 0$.}

\subsection{The chiral and electroweak breaking.}
\label{subsec:breaking}

While the electroweak symmetry is only spontaneously broken by the Higgs
boson $H = {\Bbb S}^0 ({\Bbb D}_1)$ (see the remark at the end of the
paragraph \ref{subsubsec:N=6})
getting a non vanishing vacuum expectation value $\la H \ra =
v/\sqrt{2}$, the chiral $SU(2)_L \times SU(2)_R$ symmetry is both
explicitly broken by $g' \not = 0$ and spontaneously by $\la H\ra \not =
0$.

The electroweak symmetry is broken down to the electromagnetic $U(1)_{em}$,
and the chiral symmetry down to its diagonal subgroup, the ``custodial''
$SU(2)_V$. We have indeed seen that all quadruplets decompose into a triplet
plus a singlet of $SU(2)_V$, and that the Higgs is precisely a singlet.

The electromagnetic $U(1)_{em}$ is a subgroup of $SU(2)_V$, as will be
studied in detail in the next subsection, and {\em the electroweak
spontaneous breaking is identical to the chiral breaking.}

This has consequences on the nature of the Goldstone bosons, since there are
only three of them, which become the longitudinal components of the massive
gauge fields. They are the pseudoscalar triplet $\vec {\Bbb P}({\Bbb D}_1)$.
This means in particular that:\l
- they are not aligned with any ``strong'' eigenstate (pion, kaon
$\ldots$), but they are linear combinations of them;\l
- that the strong (or flavour) eigenstates are experimentally massive (and
can be very massive) is no longer a contradiction with the spontaneous
breaking of chiral symmetry; the pion triplet would in particular only be
a triplet of Goldstone bosons if there was only one generation (meaning of
course that the mixing angles do not exist);\l
- the two ``scales'' of spontaneous breaking are identical to the mass of
the $W$'s; since in the real case of three generations, the Goldstones
``include'' the quark top, this explains why these two phenomenological mass
scales are not very different; the scale of the top quark appears as
``normal'' (with the restriction mentioned in subsection \ref{subsec:remark}).

The traditional picture of chiral symmetry breaking \cite{CurrentAlgebra}
is altered
since the relevant breaking is now that of $SU(2)_L \times SU(2)_R$ down to
$SU(2)_V$ and not that of $U(N)_L \times U(N)_R$ into the diagonal $U(N)$; 
the $U(N)_L \times U(N)_R$ chiral symmetry is {\em explicitly} broken by the 
mass terms that are introduced in a $SU(2)_L \times U(1)$ and 
$SU(2)_L \times SU(2)_R$ invariant way and the $N^2$ pseudoscalar $J=0$ mesons
do not play anymore the role of Goldstone bosons.

After the breaking of this last symmetry, there exists {\em a priori} 
two different mass scales for each multiplet, and the total number of 
(arbitrary) mass scales has doubled from $N^2/2$ to $N^2$.
In the hypothesis when the eigenstates can be split
into scalars and pseudoscalars, this means a scalar-pseudoscalar splitting
within each $({\Bbb S}^0, \vec{\Bbb P})$ or  $({\Bbb P}^0, \vec{\Bbb S})$
quadruplet. {\em One expects, as  observed, a different spectrum for scalars
and pseudoscalars}.

\subsection{The custodial $\mathbf{SU(2)_V}$ symmetry.}
\label{subsec:custodial}

I demonstrate explicitly that the present theory has a ``custodial''
$SU(2)_V$ symmetry; it is a global symmetry, which becomes local when $g'
\rightarrow 0$. A local vectorial symmetry having no anomaly can be preserved
at the quantum level.

This symmetry is {\em not} the strong isospin symmetry, because, in
particular, of the mixing angles; this means that large violation of the
strong isospin symmetry can be expected due to electroweak interactions:
the masses of mesons occurring in internal lines of the relevant diagrams
can be very different, and are likely to provide very different decay
rates for apparently similar decays if  only the isospin symmetry is
considered  (like $K \rightarrow 2\pi$ decays and the $\Delta I = 1/2$
rule).  This is currently under investigation.

One has always to keep in mind that all perturbative calculations are now to
be done with internal lines which are the $J=0$ mesons and not the quarks;
this means a different ``filter'' with which experimental data are to be
analyzed. The suggestion is consequently that the custodial symmetry might
then be found unbroken, as suggested by the extreme precision with which the
electric charge is quantized (see the next subsection). The very small
deviation found from the value $1$ for the $\rho$ parameter could
very well be due to the fact that the data have been analyzed  and computations
done up to now with a theory where fields (quarks) are not particles, and
could disappear with a new analysis. Of course, a really good one
would require to have a field theory for (at least) all mesons of
arbitrary spin, which is far from being achieved here 
\footnote{The $J=1$ mesons can be straightforwardly included in this
framework: the quadruplets now split into one triplet and one singlet of the
left, right, and diagonal $SU(2)$ groups.}
.

Another consequence concerns the ``screening'' theorem \cite{Veltman}: as
the Higgs mass can be made arbitrary without breaking the custodial
symmetry, the decoupling becomes exact at the limit where this symmetry 
is unbroken.

The 4-dimensional representations (\ref{eq:quad}) of $SU(2)_L \times U(1)$
have already been mentioned to be representations of $SU(2)_R$. They are
thus naturally representations of the diagonal $SU(2)_V$, that we study in
more detail.

When acting in the 4-dimensional vector space of which (\ref{eq:quad}) form
a basis, its generators ${\Bbb T}^3, {\Bbb T}^\pm$ can be represented
as $4\times 4$ matrices $\tilde T^3, \tilde T^\pm$; explicitly:
\begin{equation}
\tilde{\Bbb T}^+ = \left( \ba{cccc}
                   0  &  0        &  0       &  0  \cr
                   0  &  0        & \sqrt{2} &  0  \cr
                   0  &  0        &  0       &  0  \cr
                   0  & -\sqrt{2} &  0       &  0
\ea\right),\quad
\tilde{\Bbb T}^- = \left( \ba{cccc}
                   0  &  0        &  0       &  0   \cr
                   0  &  0        &  0       & -\sqrt{2} \cr
                   0  &  \sqrt{2} &  0       &  0 \cr
                   0  &  0        &  0       &  0
\ea\right),\quad
\tilde{\Bbb T}^3 = \left( \ba{cccc}
                   0  &  0        &  0       &  0  \cr
                   0  &  0        &  0       &  0  \cr
                   0  &  0        & -1       &  0  \cr
                   0  &  0        &  0       &  1
\ea\right).
\label{eq:SU2V}\end{equation}
That the first line in any of the three above matrices identically vanishes
is the translation of the already mentioned fact that the first entry
${\Bbb M}^0$ of the representations (\ref{eq:quad}) are singlets by the
diagonal $SU(2)$, while the three other entries $\vec{\Bbb M}$ form a
triplet in the adjoint representation.

The global $SU(2)_V$ symmetry occurs when the gauge fields
$W_\mu^\pm$ and $ \tilde Z_\mu = Z_\mu/\cos\theta_W$,
with $\theta_W$ the Weinberg angle,
transform like a vector in the adjoint representation of $SU(2)_V$. This is
not a surprise since those precisely absorb the $\vec{\Bbb P}({\Bbb D}_1)$ 
triplet of eq.~(\ref{eq:SP}), also in
the adjoint, to become massive, when the gauge symmetry is broken down from
$SU(2)_L \times U(1)_{\Bbb Y}$ to $U(1)_{em}$. The normalization of the last
one ensures that the resulting mass term for the gauge fields
$M_W^2 (2 W_\mu^+ W^{\mu -} + Z_\mu Z^\mu/c_W^2)$ satisfies $\rho =1$, where
$\rho = M_W/(M_Z \cos\theta_W)$.  We recover the usual link between the
custodial $SU(2)_V$ and the value of $\rho$ \cite{Sikivie}.

More precisely, we consider the Lagrangian built with covariant with respect to
$SU(2)_L \times U(1)$ derivatives from the special combination of invariant
(\ref{eq:Idiag})
\begin{equation}
\hskip -5mm
{\cal L}=
{1\over 4}
\left((\sum_{symmetric\ {\Bbb D}} - \sum_{skew-symmetric\ {\Bbb D}})
\left(D_\mu\Phi_{even}({\Bbb D})\otimes D^\mu\Phi^\dagger_{odd}({\Bbb D})
  +D_\mu\Phi_{odd}({\Bbb D})\otimes D^\mu\Phi^\dagger_{even}({\Bbb D})
\right) \right)
\label{eq:lag}
\end{equation}
The potential, being trivially invariant by $SU(2)_V$ from what has been
said in the construction of the invariants, has been omitted.

Let us explicitly write the covariant
(with respect to $SU(2)_L \times U(1)$) derivatives of a quadruplet, and
show that they transform like a singlet plus a triplet by the custodial
$SU(2)$. We do it explicitly for a $\cal P$-even quadruplet.

\vbox{
\bea
D_\mu {\Bbb M}^0_{even} &=& \p_\mu {\Bbb M}^0_{even} + {e\over 2 s_W}
                   (W_\mu^1{\Bbb M}^1_{even}  + W_\mu^2{\Bbb M}^2_{even}
                    + (Z_\mu/c_W){\Bbb M}^3_{even} ),\cr
     &=& {\cal D}_\mu {\Bbb M}^0_{even} + {e\over 2 s_W}
                  (W_\mu^1{\Bbb M}^1_{even}  + W_\mu^2{\Bbb M}^2_{even}
                    + (Z_\mu/c_W){\Bbb M}^3_{even} ),\cr
D_\mu {\Bbb M}^3_{even} &=& \p_\mu {\Bbb M}^3_{even} + {e\over 2 s_W} \left(
                   i\,(W_\mu ^+ {\Bbb M}^-_{even}  - W_\mu^-{\Bbb
M}^+_{even})
                      - (Z_\mu/c_W){\Bbb M}^0_{even} \right),\cr
     &=& {\cal D}_\mu {\Bbb M}^3_{even}
                        -{e\over 2 s_W}(Z_\mu /c_W){\Bbb M}^0_{even},\cr
D_\mu {\Bbb M}^+_{even} &=& \p_\mu {\Bbb M}^+_{even} -{e\over 2 s_W} \left(
                    W_\mu^+({\Bbb M}^0_{even} + i {\Bbb M}^3_{even})
             -i (Z_\mu/c_W){\Bbb M}^+_{even}\right)
             +i {e\over c_W}B_\mu {\Bbb M}^+_{even}, \cr
     &=& {\cal D}_\mu {\Bbb M}^+_{even}
                 -{e\over 2 s_W} W_\mu^+{\Bbb M}^0_{even}
                 +i{e\over c_W} B_\mu {\Bbb M}^+_{even},\cr
D_\mu {\Bbb M}^-_{even} &=& \p_\mu {\Bbb M}^-_{even} -{e\over 2 s_W} \left(
                    W_\mu^-({\Bbb M}^0_{even} - i {\Bbb M}^3_{even})
             +i (Z_\mu/c_W){\Bbb M}^-_{even}\right)
             -i {e\over c_W} B_\mu {\Bbb M}^-,\cr
     &=& {\cal D}_\mu {\Bbb M}^-_{even}
                  -{e\over 2 s_W} W_\mu^-{\Bbb M}^0_{even}
                   -i{e\over c_W} B_\mu {\Bbb M}^-_{even}.
\label{eq:covders}
\eea
}
In eq.~({\ref{eq:covders}) above, we noted $c_W$ and $s_W$ respectively the
cosine and sine of the Weinberg angle.  $A_\mu$ is the photon, $W_\mu^\pm
= (W_\mu^1 \pm i W_\mu^2) /\sqrt{2}$, and we have as usual
\bea
g &=& {e\over s_W},\  g' = {e\over c_W},\cr
Z_\mu &=& c_W W_\mu^3 - s_W B_\mu,\  A_\mu = c_W B_\mu + s_W W_\mu^3.
\eea
${\cal D}_\mu$ is the covariant derivative with respect to the diagonal
$SU(2)_V$ group
\begin{equation}
{\cal D}_\mu {\Bbb M} = \p_\mu {\Bbb M}
            -i{e\over s_W}({1\over\sqrt{2}}(W_\mu^+ \tilde{\Bbb T}^-
       + W_\mu^- \tilde{\Bbb T}^+)
+{Z_\mu \over c_W}\tilde {\Bbb T}^3).{\Bbb M}\ .
\end{equation}
The normal derivative of $\Bbb M$ transforming like $\Bbb M$ itself,
that $D_\mu {\Bbb M}^0$ is a singlet of $SU(2)_V$ is trivial as soon as, as
stressed before, $\vec{\Bbb M}$ is a triplet in the adjoint and $(W_\mu
^\pm,
Z_\mu/c_W)$ too, since the scalar product of those two vectors is an
invariant;
\l
that the three other covariant derivatives transform like a vector results
from the three following facts:\l
- from the 2 vectors $\vec{\Bbb M}$ and $(W_\mu ^\pm,Z_\mu/c_W)$ we can form
a third one with the $\epsilon_{ijk}$ tensor
\begin{equation} \left(\ba{l}
{\Bbb M}^- W_\mu^+    - {\Bbb M}^+ W_\mu^-,\cr
{\Bbb M}^3 W_\mu^+    - {\Bbb M}^+ (Z_\mu/c_W),\cr
{\Bbb M}^3 W_\mu^-    - {\Bbb M}^- (Z_\mu/c_W);
\ea\right)\end{equation}
- ${\Bbb M}^0$ being a singlet by $SU(2)_V$, the terms ${\Bbb M}^0
W_\mu^\pm$
transform like $W_\mu^\pm$ and thus like ${\Bbb M}^\pm$, $(Z_\mu /c_W){\Bbb
M}^0$ like $(Z_\mu /c_W)$ and thus like ${\Bbb M}^3$;\l
- $B_\mu$ is to be considered as a singlet of $SU(2)_V$, such that the terms
$(B_\mu /c_W){\Bbb M}^\pm$ transform like ${\Bbb M}^\pm$.

The same argumentation works for $\cal P$-odd scalars. Their covariant
derivatives are immediately obtained from eqs.~(\ref{eq:covders}) above by
changing the signs of all ${\Bbb M}^0$'s.

This shows the existence of a  global $SU(2)_V$ custodial symmetry for
the Lagrangian,
independently of the value of the hypercharge coupling $g'$.

Let us now examine whether this symmetry can be considered as a
local symmetry.

Making a space-time dependent $SU(2)_V$ transformation with parameters
$\vec\theta$ on the scalar fields and transforming the vector fields
$W_\mu^\pm, Z_\mu/c_W$ like the corresponding gauge potentials
($B_\mu$ being a singlet does not transform),
ones finds from (\ref{eq:covders}) that the Lagrangian
(\ref{eq:lag}) varies, for each quadruplet, by
\begin{equation}
\Delta{\cal L} = {\cal D}_\mu\vec\theta .(\vec{\Bbb M}\otimes D^\mu {\Bbb
M}^0
                  -{\Bbb M}^0\otimes  D^\mu\vec{\Bbb M}),
\label{eq:deltaL}\end{equation}
such that the existence of a  local custodial $SU(2)_V$ symmetry is
linked to the conservation of the triplet of currents $\vec V^\mu$
\begin{equation}
{\cal D}_\mu \vec V^\mu = 0,
\end{equation}
with
\begin{equation}
\vec V^\mu = \vec {\Bbb M}\otimes  D^\mu {\Bbb M}^0
              - {\Bbb M}^0\otimes  D^\mu \vec{\Bbb M}.
\label{eq:vmu}\end{equation}
$\vec V_\mu$ is an $SU(2)_V$ triplet. Its ``singlet'' partner $V_\mu^0$
identically vanishes by the definition (\ref{eq:vmu}).

These currents are automatically covariantly (with respect to $SU(2)_L
\times U(1)$) conserved by the classical
equations of motion for the $\Bbb M$ fields, as can be seen from
(\ref{eq:vmu}), which entails
\begin{equation}
D^\mu V^i_\mu = {\Bbb M}^i\otimes  D^2 {\Bbb M}^0 -
                                     {\Bbb M}^0\otimes  D^2 {\Bbb M}^i,
\end{equation}
and  from the Lagrangian (\ref{eq:lag}) to which we can add any term
quadratic in the invariants $\cal I$ for any quadruplet.

Now,
\begin{equation}
D^\mu V^i_\mu = {\cal D}^\mu V^i_\mu -ig' B_\mu \tilde{\Bbb Q}.V^i_\mu,
\end{equation}
where we have used the Gell-Mann-Nishijima relation and the fact that,
since $V_\mu ^0$ identically vanishes, the ``left'' $SU(2)_L$ acts on $\vec
V_\mu$ like the diagonal $SU(2)_V$.

We can thus conclude that the custodial symmetry, which is a  global
symmetry, becomes  local when the hypercharge coupling $g'$ goes to zero.

Anomaly-free, it is thus an exact  local
symmetry of the standard $SU(2)_L \times U(1)$ Lagrangian
(\ref{eq:lag}) for $J=0$ fields, with gauge fields $W_\mu^\pm,
Z_\mu/c_W$.

\subsection{Quantization of the electric charge; electric magnetic duality.}
\label{subsec:duality}

The two known ways to explain the quantization of the electric charge are
\cite{Olive}:\l
- that the corresponding generator is the ``{\em z}'' component of an
angular momentum ($SU(2)$);\l
- that there exists at least one magnetic ``monopole''-like object (Dirac
quantization).

The idea of electric-magnetic duality is the those two mechanisms are just
two aspects of the same phenomenon and always occur simultaneously.

It is thus suggestive that, at the same time I showed that
in the standard model for $J=0$ mesons the electric charge generator is
precisely the ``{\em z}'' component of the custodial $SU(2)_V$, Cho and
Maison \cite{ChoMaison} showed that the scalar sector of the standard
model has, because of the presence of the hypercharge $U(1)$, the right
topological structure ($CP^1$) to incorporate dyon-like solutions, 
which they exhibited numerically. The problem of the infinite zero-point 
energy of their solutions was later shown \cite{ChoKimm} to be regularized 
when the group of symmetry is slightly enlarged and/or new interactions
introduced. The fact that the model presented here also
incorporates a chiral $SU(2)_L \times SU(2)_R$ symmetry might also help
regularizing their classical solutions.

The model proposed here seems  consequently to present
the right properties to achieve electric-magnetic duality.

The interest of such a property lies also in the fact that one then expects a
strongly interacting sector, in which the fields are monopole-like extended
objects. Those skyrmion-like particles \cite{Skyrme}, built ``on top of''
mesons, are then natural candidates for baryons.

Starting from an electroweak model of physical particles, we reach the idea
that at least a certain aspect of the strong interactions could be included
as another sector of the theory (strong interactions of mesons can originate
from the high mass limit of the Higgs boson \cite{LeeQuiggThacker}),
towards a true unification of non-gravitational interactions.

\section{$\mathbf{CP}$ violation.}
\label{section:CP}

All phenomena of $CP$ violation
\cite{ChristensonCroninFitchTurlay}\cite{Argus} are, up to now, compatible with
the so-called ``indirect'' violation \cite{Nir}, explained by the  electroweak
mass eigenstates not being $CP$ eigenstates. The stakes are high for the
observation of ``direct'' $CP$ violation, and, in particular, for
discovering whether the so-called $\epsilon '$ parameter is vanishing or not.

I show below, that, in the present framework, unitarity requires
that the electroweak mass eigenstates are always $C$ eigenstates and that
``indirect'' $CP$ violation only occurs as a consequence of $P$ violation.

Furthermore, I show, and this is an immediate an very simple consequence of
the nature of the $J=0$ electroweak representations constructed above,
that, even when there is a complex phase in the mixing matrix $\Bbb K$,
electroweak mass eigenstates can still be $CP$ eigenstates.
So, the existence of a complex mixing matrix at the fermionic level is no
longer a sufficient condition for mesonic electroweak mass eigenstates to be
different from $CP$ eigenstates.

``Indirect'' $CP$ violation consequently fades away, and the true search for
$CP$ violation should really be concentrated on that of ``direct'' $CP$
violation. Phrased in a more provocative way, it seems that we do not know
yet if $CP$ is truly violated.

\subsection{Electroweak versus $\mathbf{CP}$ eigenstates.}
\label{subsec:eigenstates}

The electroweak Lagrangian for $J=0$ mesons is the one  of
eq.~(\ref{eq:lag}) plus the potential built from quadratic
invariants according to section \ref{section:lagrangian}.

Unitarity compels this Lagrangian to be hermitian, in particular its
quadratic part.

Its diagonalization yields the electroweak mass eigenstates.
Let us restrict for the sake of
simplicity to a subsystem of two non-degenerate electroweak mass eigenstates
$U$ and $V$; they are in general complex linear combinations of quadruplets
(\ref{eq:SP}) and (\ref{eq:PS}), and transform by $SU(2)_L$
according to (\ref{eq:actioneven}). $\cal L$ writes, for example
\begin{equation}
{\cal L} = {1\over 2}(\p_\mu U\otimes \p^\mu U -\p_\mu V\otimes \p^\mu V
 - m_U^2 U\otimes U + m_V^2 V\otimes V +\cdots).
\end{equation}
with $m_U^2 \not = m_V^2$, where wee have only written above the quadratic part.

Hermiticity  yields the two following  equations,
coming respectively from the kinetic and mass terms
\begin{equation}\left\{\ba{l}
(U\otimes U - V\otimes V)^\dagger = U\otimes U - V\otimes V,\cr
(m_U^2 U\otimes U - m_V^2 V\otimes V)^\dagger =m_U^2 U\otimes U - m_V^2
V\otimes V,
\ea\right. \end{equation}
which, if we reject complex values of the (mass)$^2$, entail
\begin{equation} U = \pm U^\dagger,\quad V=\pm V^\dagger;
\end{equation}
unitarity thus requires  that the electroweak mass eigenstates be also
$C$ eigenstates.

Consequence: {\em if electroweak mass eigenstates are observed not to be $CP$
eigenstates, they can only be  mixtures of states with different parities.}

We had already mentioned that the most general eigenstates in this $P$
violating theory do not have, as expected, a definite parity.
This transforms the problem of indirect $CP$ violation into finding an
explanation for the {\em smallness of the observed mixture between scalars
and pseudoscalars}.

\subsection{The fading role of the Kobayashi-Maskawa mixing matrix.}
\label{subsec:KM}

Suppose that we have a complex mixing matrix $\Bbb K$; the following
Lagrangian for $J=0$ mesons, where the sum is extended to all
representations defined by eqs.~(\ref{eq:SP},\ref{eq:PS},\ref{eq:quad}),
is nevertheless hermitian,
($D_\mu$ is the covariant derivative with respect to $SU(2)_L \times U(1)$)

\vbox{
\bea
{\cal L}= &&{1\over 2}\sum_{symmetric\ {\Bbb D}}
           \left(D_\mu ({\Bbb S}^0, \vec {\Bbb P})(\Bbb D)
                             \otimes D^\mu ({\Bbb S}^0, \vec {\Bbb P})(\Bbb
D)
          - m_D^2 ({\Bbb S}^0, \vec {\Bbb P})(\Bbb D)
                             \otimes ({\Bbb S}^0, \vec {\Bbb P})(\Bbb D)
        \right.\cr
&&\hphantom{{1\over 2}\sum_{symmetric\ {\Bbb D}}}
\left. -\left(D_\mu ({\Bbb P}^0, \vec {\Bbb S})(\Bbb D)
                    \otimes D^\mu ({\Bbb P}^0, \vec {\Bbb S})(\Bbb D)
          - \ti m_D^2 ({\Bbb P}^0, \vec {\Bbb S})(\Bbb D)
                    \otimes ({\Bbb P}^0, \vec {\Bbb S})(\Bbb D)
\right) \right)\cr
- &&{1\over 2}\sum_{skew-symmetric\ {\Bbb D}}
\left(D_\mu ({\Bbb S}^0, \vec {\Bbb P})(\Bbb D)
                     \otimes D^\mu ({\Bbb S}^0, \vec {\Bbb P})(\Bbb D)
          - m_D^2 ({\Bbb S}^0, \vec {\Bbb P})(\Bbb D)
                     \otimes ({\Bbb S}^0, \vec {\Bbb P})(\Bbb D) \right.\cr
&&\hphantom{{1\over 2}\sum_{symmetric\ {\Bbb D}}}
        \left.-\left(D_\mu ({\Bbb P}^0, \vec {\Bbb S})(\Bbb D)
                     \otimes D^\mu ({\Bbb P}^0, \vec {\Bbb S})(\Bbb D)
          - \ti m_D^2 ({\Bbb P}^0, \vec {\Bbb S})(\Bbb D)
                \otimes ({\Bbb P}^0, \vec {\Bbb S})(\Bbb
D)\right)\right),\cr
& &
\eea
}
and its mass eigenstates, being the $({\Bbb S}^0, \vec {\Bbb P})$ and
$({\Bbb P}^0, \vec {\Bbb S})$ representations given by
(\ref{eq:SP},\ref{eq:PS}), are $CP$ eigenstates \cite{Machet1}.
It is of course straightforward to also build hermitian $SU(2)_L \times
U(1)$ invariant quartic terms.

Consequence: {\em The existence of a complex phase in the mixing matrix for
quarks is not a sufficient condition for the existence of electroweak mass
eigenstates for $J=0$ mesons different from $CP$ eigenstates}.

We have indeed seen that, at the mesonic level, all dependence on the mixing
matrix $\Bbb K$ can be reabsorbed in the definition of the asymptotic
states.

\section{The link with observed mesons.}
\label{section:mesons}

It is  shown below that the fields that we have been
dealing with are in one-to-one correspondence with the observed scalar and
pseudoscalar $J=0$ mesons.

For this purpose we shall study in particular their leptonic decays.
After general considerations about which kind are expected to decay into
leptons,  are which are not, we show that one recovers the standard PCAC
result by a simple rescaling of the fields and couplings: the scaling
parameter is 
\begin{equation}
a = {f\over\la H\ra},
\label{eq:scaling}
\end{equation}
where $f$ is the leptonic decay constant, supposed here to be the same for
all mesons.

We make some general remarks about semi-leptonic decays.

All computations are made at tree-level, with the propagators of the
massive gauge bosons taken in the unitary gauge.

\subsection{General selection rules.}
\label{subsec:rules}

The leptonic and semi-leptonic decays  of $J=0$ mesons occur via the crossed
terms in the kinetic terms of the Lagrangian (\ref{eq:lag}) which are 
proportional to
\begin{equation}
\p_\mu {\Bbb M} \otimes g\vec W^\mu \vec {\Bbb T}. {\Bbb M},
\end{equation}
where I have used a shortened and symbolic notation in which the $\Bbb M$'s
are  the ingoing and outgoing (if any) meson, $g$ one of the two 
$SU(2)_L \times U(1)$ coupling constants,
$\vec W^\mu$ the set of three massive gauge fields (one of them can be the $Z$).
The gauge field then couples to the two outgoing leptons.

If ${\Bbb T}.{\Bbb M}$ yields the Higgs boson, then the term proportional
to $\la H \ra$ triggers a leptonic decay, like described in fig.~1; 

\vbox{
\figskip
\begin{center}
\epsfig{file=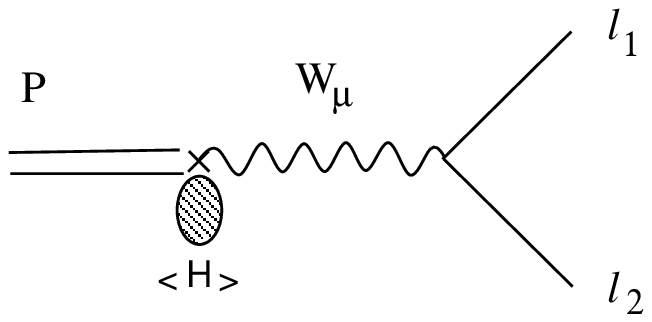}
\end{center}
\centerline{\em Fig.~1: The leptonic decay of a pseudoscalar meson.}
\figskip
}

if it yields another meson, then the process is a semi-leptonic decay,
as described in fig.~2.

\vbox{
\figskip
\begin{center}
\epsfig{file=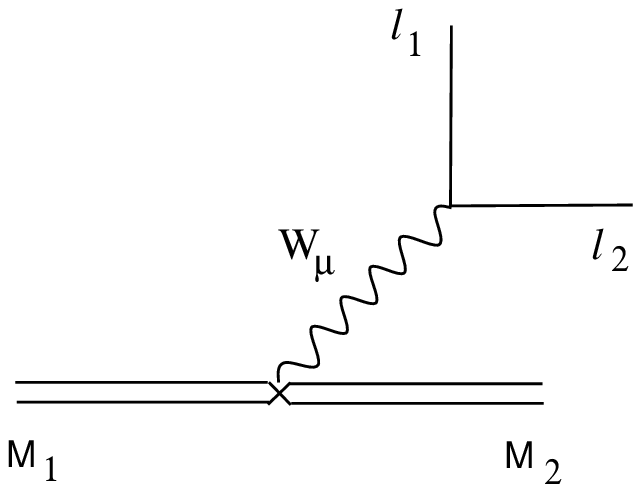}
\end{center}
\centerline{\em Fig.~2: The semi-leptonic decay of a meson.}
\figskip
}

The above mechanisms have immediate and simple consequences, and ``selection
rules'' result:\l
- only mesons which have a non-vanishing projection on the Higgs boson when
acted upon by one of the generators of the electroweak group can decay
into leptons;\l
- hence, if we suppose that the Higgs is unique and is a pure (neutral) scalar 
(which is not {\em a priori} true in a parity violating theory),
then {\em scalar mesons never decay in a pure leptonic way}:
indeed, when acted
upon by a generator of the group they can only give a scalar or a
pseudoscalar; if it is a scalar, it can only be, as can be seen from
eq.~(\ref{eq:actioneven}) a charged one, thus different from the Higgs boson
and which does not condensate in the vacuum; if it is a pseudoscalar, 
it does not condense either;\l
- all scalars and pseudoscalars can decay semi-leptonically; however, when
acted upon by a generator of the electroweak group, any $SU(2)_V$ (neutral)
singlet in the representations (\ref{eq:SP},\ref{eq:PS}) is transmuted
into a particle with opposite parity; a scalar $SU(2)_V$ singlet will
consequently only semi-leptonically decay into a pseudoscalar and {\em vice
versa}; also, the neutral of the $SU(2)_V$ triplet can only give the singlet,
with opposite parity, when acted upon by ${\Bbb T}^3_L$; as a consequence,
the neutral semi-leptonic decays of a neutral $SU(2)_V$ triplet always gives
a neutral outgoing particle with opposite parity. We can thus state the
rule: {\em the neutral particle produced by the semi-leptonic decay of an
incoming neutral $J=0$ meson has always a parity opposite to that of the
incoming particle}.

An immediate consequence is that scalars are difficult to detect since they
do not have leptonic decays; so are consequently semi-leptonic decays of
neutral pseudoscalar mesons which yield scalar neutral mesons.

Decays of neutral mesons are furthermore severely constrained by the absence
of flavour changing neutral currents: it can indeed be checked (this is
easily understood since the present model has been built in a way
compatible with the Glashow-Salam-Weinberg model) that
this selection rule is still valid here: no decay is allowed that would
require flavour changing neutral currents at the fermionic level.
In practice, the semi-leptonic decay of a neutral meson can only yield
another neutral meson when the decaying particle is ``diagonal in
flavour''.

Four outgoing leptons can originate from a neutral scalar decaying
semi-leptonically: two leptons come from the gauge field and the two others
from the leptonic decay of the produced pseudoscalar.

\subsection{Explicit representations for $\mathbf{N=4}$.}
\label{subsec:explicit}

For the sake of simplicity, we shall work in this section in the case of
two generations $N=4$.

The four types of $SU(2)_L \times U(1)$ quadruplets that now arise,
corresponding respectively to the matrices ${\Bbb D}_i, i=1 \cdots 4$ of
subsection \ref{subsec:special} are:

\vbox{
\bea
& &\Phi({\Bbb D}_1) = \cr
& & \hskip -2cm \left[
{1\over\sqrt{2}} \left(\ba{rrcrr} 1 &   &\vline &    &    \nonumber\\
                                    & 1 &\vline &    &    \nonumber\\
                                    \hline
                                    &   &\vline &  1 &    \nonumber\\
                                    &   &\vline &    &  1 \ea \right),
{i\over\sqrt{2}} \left(\ba{rrcrr} 1 &   &\vline &   &   \nonumber\\
                                    & 1 &\vline &   &   \nonumber\\
                                    \hline
                                    &   &\vline &-1 &   \nonumber\\
                                    &   &\vline &   & -1   \ea \right),
i \left(\ba{rrcrr}   &  &\vline & c_\theta &  s_\theta \nonumber\\
                             &  &\vline &-s_\theta &  c_\theta \nonumber\\
                            \hline
                             &  &\vline &   &     \nonumber\\
                             &  &\vline &   &  \ea \right),
i \left(\ba{rrcrr}   &   &\vline &   &   \nonumber\\
                             &   &\vline &   &   \nonumber\\
                             \hline
                             c_\theta &-s_\theta &\vline &   &
\nonumber\\
                             s_\theta & c_\theta &\vline &   &   \ea \right)
\right]; \nonumber\\
& &
\label{eq:PHI1}
\eea
}
\vbox{
\bea
& &\Phi({\Bbb D}_2) = \cr
& & \hskip -2cm \left[
{1\over\sqrt{2}} \left(\ba{rrcrr}
     1 &   &\vline &    &    \nonumber\\
       & -1 &\vline &    &    \nonumber\\
     \hline
  &  & \vline & c_\theta^2 - s_\theta^2 & 2c_\theta s_\theta  \nonumber\\
  &  & \vline & 2c_\theta s_\theta & s_\theta^2 -c_\theta^2  \ea \right),
{i\over\sqrt{2}} \left(\ba{rrcrr}
             1 &    & \vline &   &   \nonumber\\
               & -1 & \vline &   &   \nonumber\\
             \hline
      &  & \vline & s_\theta^2 - c_\theta^2 & -2c_\theta s_\theta
\nonumber\\
      &  & \vline & -2c_\theta s_\theta & c_\theta^2 - s_\theta^2  \ea
\right),
\right .\nonumber\\
& & \hskip 7cm \left .
i \left(\ba{rrcrr}   &  &\vline & c_\theta & s_\theta \nonumber\\
                             &  &\vline & s_\theta & -c_\theta \nonumber\\
                            \hline
                             &  &\vline &   &     \nonumber\\
                             &  &\vline &   &  \ea \right),
i \left(\ba{rrcrr}     &   &\vline &   &   \nonumber\\
                               &   &\vline &   &   \nonumber\\
                              \hline
                              c_\theta & s_\theta &\vline &   & \nonumber\\
                              s_\theta & -c_\theta &\vline &   &   \ea
\right)
\right]; \nonumber\\
& &
\label{eq:PHI2}
\eea
}
\vbox{
\bea
& &\Phi({\Bbb D}_3) = \cr
& & \hskip -2cm \left[
{1\over\sqrt{2}} \left(\ba{rrcrr}
        & 1 &\vline &    &    \nonumber\\
      1 &  &\vline &    &    \nonumber\\
     \hline
     &  & \vline & -2c_\theta s_\theta & c_\theta^2 - s_\theta^2
\nonumber\\
     &  & \vline &  c_\theta^2 - s_\theta^2 & 2c_\theta s_\theta   \ea
\right),
{i\over\sqrt{2}} \left(\ba{rrcrr}
         & 1 & \vline &   &   \nonumber\\
       1 &  & \vline &   &   \nonumber\\
       \hline
     &  & \vline & 2c_\theta s_\theta & s_\theta^2 - c_\theta^2 \nonumber\\
     &  & \vline & s_\theta^2 - c_\theta^2 & -2c_\theta s_\theta \ea
\right),
\right .\nonumber\\
& & \hskip 7cm \left .
i \left(\ba{rrcrr}   &  & \vline & -s_\theta & c_\theta \nonumber\\
                             &  & \vline & c_\theta &  s_\theta \nonumber\\
                            \hline
                            &  &\vline &   &     \nonumber\\
                            &  &\vline &   &  \ea \right),
i \left(\ba{rrcrr}     &   &\vline &   &   \nonumber\\
                               &   &\vline &   &   \nonumber\\
                              \hline
                             -s_\theta & c_\theta &\vline &   &
\nonumber\\
                              c_\theta & s_\theta &\vline &   &   \ea
\right)
\right]; \nonumber\\
& &
\label{eq:PHI3}
\eea
}
\vbox{
\bea
& &\Phi({\Bbb D}_4) = \cr
& & \hskip -2cm \left[
{1\over\sqrt{2}} \left(\ba{rrcrr}           & 1 &\vline &    & \nonumber\\
                                         -1 &   &\vline &    & \nonumber\\
                                            \hline
                                            &   &\vline &    & 1
\nonumber\\
                                            &   &\vline & -1 &    \ea
\right),
{i\over\sqrt{2}}\left(\ba{rrcrr}   & 1 &\vline &   &   \nonumber\\
                                         -1 &   &\vline &   &   \nonumber\\
                                           \hline
                                            &   &\vline &   & -1
\nonumber\\
                                            &   &\vline & 1 &    \ea
\right),
 i \left(\ba{rrcrr}   &  &\vline & -s_\theta & c_\theta \nonumber\\
                            &  &\vline & -c_\theta & -s_\theta \nonumber\\
                            \hline
                            &  &\vline &   &     \nonumber\\
                            &  &\vline &   &  \ea \right),
i  \left(\ba{rrcrr}     &   &\vline &   &   \nonumber\\
                              &   &\vline &   &   \nonumber\\
                            \hline
                           s_\theta & c_\theta &\vline &   &   \nonumber\\
                           -c_\theta & s_\theta &\vline &   &   \ea \right)
\right]; \nonumber\\
& &
\label{eq:PHI4}
\eea
}
%
$c_\theta$ and $s_\theta$ stand respectively for the cosine and sine of the
Cabibbo angle $\theta_c$.

We shall also use in the following the notations
\begin{equation}
({\Bbb S}^0, \vec{\Bbb P})({\Bbb D}_1)= \Phi_1,\quad
({\Bbb S}^0, \vec{\Bbb P})({\Bbb D}_2)= \Phi_2,\quad
({\Bbb S}^0, \vec{\Bbb P})({\Bbb D}_3)= \Phi_3,\quad
({\Bbb S}^0, \vec{\Bbb P})({\Bbb D}_4)= \Phi_4.
\label{eq:notation}\end{equation}
According to the remark of subsubsection \ref{subsubsec:N=6}, we consider the
Higgs boson to be the unique scalar singlet with a non vanishing trace
\begin{equation}
H = {\Bbb S}^0 ({\Bbb D}_1).
\label{eq:Higgs}
\end{equation}
{}From the general selection rules written above, only the pseudoscalar mesons
which have a non-vanishing projection on the three Goldstones
$\vec{\Bbb P}({\Bbb D}_1)$ will undergo leptonic decays.

\subsection{From matrix-fields to observed mesons: the case of leptonic
decays.}
\label{subsec:leptonic}

We call $\Psi_\ell$ the leptonic equivalent of $\Psi$ in
eq.~(\ref{eq:Psi}).

Let us rescale the fields according to:\l
- for mesons and leptons:
\begin{equation}
\Phi = a \Phi',\quad
\Psi_\ell = a \Psi'_\ell;
\end{equation}
in particular one has $H = a H'$;\l
- for the  gauge fields, generically noted $\sigma_\mu$:
\begin{equation}
\sigma_\mu = a \sigma'_\mu;
\end{equation}
and  all coupling constants, called generically $\kappa$, according to
\begin{equation}
\kappa = {\tilde\kappa \over a}.
\end{equation}
The fields and coupling constants to be considered as physical are the
rescaled ones. 

The relations between the matrix-valued $\Bbb M$ mesonic
fields and the physically observed ``strong'' eigenstates (kaon, pion
$\ldots$) is given by relations like eq.~(\ref{eq:Kpi}) below for ${\Bbb
P}^+({\Bbb D}_1)$, which can be read off directly from eqs.~(\ref{eq:PHI1})
to (\ref{eq:PHI4}). The translation is most easily done (see section
\ref{section:UN}) by sandwiching the
$\Bbb M$ matrices between $\Psi$ and $\bar\Psi$ to find its components on
the strong eigenstates.

The Lagrangian we furthermore rescale by $1/a^2$ in order that the kinetic
terms are normalized to ``$1$'' when expressed in terms of the ``primed''
fields.

The propagators of the gauge fields are left unchanged, in particular those
of the massive $W$'s and $Z$ since $g^2 \la H \ra ^2 = \tilde g^2 \la H' \ra
^2$.

Suppose that the incoming meson in fig.~1 is, for example a ``strong'' $K^+$,
that is, in the quark notation, a $\bar u \gamma_5 s$ state
created by strong interactions.

As it is the relevant part for leptonic decays, we only rewrite, according
to eq.~(\ref{eq:covders}), the kinetic terms for $\Phi_1$ in terms of the 
rescaled fields, like for example
\begin{equation}
{\Bbb P}^+({\Bbb D}_1) = a {\Bbb P}^{'+}({\Bbb D}_1) =
a \left(c_\theta (\pi^+ + D_s ^+) + s_\theta(K^+ + D^+)\right);
\label{eq:Kpi}
\end{equation}
One has accordingly
\bea
{\cal L}' = {1\over a^2} {\cal L} &=&
{1\over a^2}{1\over 2} \left( \p_\mu {\Bbb P}^+ - {g\over 2} W_\mu^+ H\right)
\otimes \left( \p_\mu {\Bbb P}^- - {g\over 2} W_\mu^- H\right)+\cdots\cr
&=& 
{1\over 2}\left(\p_\mu {\Bbb P}^{'+} - {\tilde g\over 2} W_\mu^{'+} H'\right)
\otimes 
   \left(\p_\mu {\Bbb P}^{'-} - {\tilde g\over 2} W_\mu^{'-} H'\right)+\cdots;
\eea
on the leptonic side the coupling of the gauge fields to the electron and
the neutrino writes ($\gamma^\mu_L = \gamma^\mu (1-\gamma_5)/2$)
\begin{equation}
{\cal L}'_\ell ={1\over a^2}{\cal L}_\ell = {1\over a^2}
(g W_\mu^+  e^- \gamma^\mu_L \bar\nu_\ell)
= \tilde g W_\mu^{'+} e^{'-} \gamma^\mu_L \bar\nu_\ell'.
\end{equation}
The diagram of fig.~1, expressed in terms of the rescaled fields and
coupling constant yields 
\bea
& &K^+\ {1\over 2}(s_\theta\; i\tilde g\; ik_\mu  \la H'\ra)
({4i\over \tilde g^2 \la H'\ra^2}) (i\tilde g)\ 
                                 e^{'-} \gamma^\mu_L \bar\nu_\ell'\cr
&=&{1\over 2} s_\theta {4k_\mu \over \la H' \ra}\ K^+ 
                  \ e^{'-} \gamma^\mu_L \nu_\ell'\cr
&=& {1\over 2}s_\theta \;f\; {4k_\mu \over \la H\ra^2}\ K^+ 
                  \ e^{'-} \gamma^\mu_L \bar\nu_\ell'
    = {1\over 2}s_\theta \;f\; k_\mu{g^2 \over M_W^2}\ K^+
                  \ e^{'-} \gamma^\mu_L \bar\nu_\ell'
\eea
which is exactly, for the rescaled (physical) fields,
the result traditionally obtained by PCAC.
We have used eq.~(\ref{eq:scaling}), the relation 
$M_W^2 = g^2 \la H \ra ^2 /4 = \tilde g^2 \la H' \ra ^2 /4$ and the fact that,
in the unitary gauge, the $W$ propagator $D^{\mu\nu}_W$ satisfies
\begin{equation}
ik_\nu D^{\mu\nu}_W(k) = -{k_\mu \over M_W^2},
\end{equation}
where $k_\mu$ is the momentum of the incoming meson.

\section{An extension to the  leptonic sector.}
\label{section:leptons}

We have now at our disposal a renormalizable gauge theory for
$J=0$ mesons which is anomaly-free, and in which the quantization of the
electric charge for asymptotic states has been correlated with a custodial
$SU(2)_V$ symmetry to stay unbroken at the quantum level.

There is now a need to also modify the leptonic sector \cite{Machet3} since:\l
- charge quantization should also hold for the corresponding asymptotic
states; if we suppose that the same mechanism 
is at work, then the theory that we are looking for should have the 
same custodial symmetry as the one unraveled above;\l
- anomalies \cite{AdlerBellJackiw} can now only spring out of fermions, such
that this sector
should be anomaly-free by itself; we cannot rely anymore on a
cancelation between quarks and leptons \cite{BouchiatIliopoulosMeyer}.

It is also well known \cite{Ramond}
that there exist problems with Weyl fermions making desirable a vector-like
theory of weak interactions.

This is why we propose to start from the purely vectorial theory
studied in \cite{BellonMachet}.
 
We shall not question universality and only deal here with one generation of
fermions.

\subsection{The custodial symmetry for a vectorial theory.}
\label{subsec:vectorial}

In the mesonic case, we have seen that each quadruplet (complex doublet) of
$SU(2)_L$ was also the sum of one $SU(2)_V$ real triplet with electric
charges $(-1,0,+1)$ plus one real chargeless singlet; this made easy and
straightforward, in the space spanned by these representations, the
connection between the custodial group of symmetry and its electromagnetic
subgroup.

Now, in the Glashow-Salam-Weinberg model \cite{GlashowSalamWeinberg}
for leptons, we do not have any 
more one complex $SU(2)_L$ doublet, but a set of doublets for left-handed
fields and singlets for right-handed ones. Implementing a custodial $SU(2)$
symmetry is consequently less intuitive here, and make us consider the
standard model for leptons as only an effective theory.

\subsubsection{Groups and representations.}

Because the notion of left and right-handed groups has a very precise
meaning when leptons are concerned, it is useful here to change
the notation and call ${\cal G}_1$ and ${\cal G}_2$ the two $SU(2)$'s
which build the equivalent of the chiral $SU(2) \times SU(2)$ group of
subsection \ref{subsec:chiralSU2}.
As we shall see later,  there however exists a similarity between ${\cal G}_1$
and the $SU(2)_L$ group of the Glashow-Salam-Weinberg model, in that they
act  in the same way on the left-handed (neutrino, electron) doublet.

Consider the quadruplet ${\cal Q}_L$ of left-handed fields
\begin{equation}
{\cal Q}_L = \left({\Bbb L}^0, {\Bbb L}^3, {\Bbb L}^+, {\Bbb L}^- \right)
         =  \left(
         -i{\nu -\nu^c \over\sqrt{2}}, {\nu + \nu^c \over\sqrt{2}},
          \ell^+, \ell^-
                                  \right)_L,
\end{equation}
$\ell^+$ and $\ell^-$, $\nu^c$ and $\nu$ are charge conjugate:
\begin{equation}
\ell^+ = C\ol{\ell^-}^T, \quad \nu^c = C{\ol\nu}^T;
\end{equation}
the superscript ``$T$'' means ``transposed'' and $C$ is the charge-conjugation
operator: $C = i\gamma_2\gamma_0$ in the Dirac representation.
The convention that $\ell^+ = (\ell^1 + i \ell^2)/\sqrt{2}$ is the charge
conjugate of $\ell^- = (\ell^1 - i \ell^2)/\sqrt{2}$ entails that
$i$ gives $-i$ by charge conjugation
and that the charge conjugate $({\cal Q}_L)^c$ of ${\cal Q}_L$ is its
right-handed counterpart ${\cal Q}_R = ({\Bbb R}^0, \vec{\Bbb R})$.

By analogy with eq.~(\ref{eq:actioneven}), we  define the actions of
${\cal G}_1$ with generators $\vec{\Bbb T}_1$ and ${\cal G}_2$
with generators $\vec{\Bbb T}_2$ on ${\cal Q}_L$ by

\vbox{
\bea
{\Bbb T}^i_1 . {\Bbb L}^j &=&
                      {i\over 2}(\epsilon_{ijk}{\Bbb L}^k +
                                \delta_{ij}{\Bbb L}^0), \cr
{\Bbb T}^i_1 . {\Bbb L}^0 &=& -{i\over 2} {\Bbb L}^i,
\label{eq:G1}\eea
}

and

\vbox{
\bea
{\Bbb T}^i_2 . {\Bbb L}^j &=&
                              {i\over 2}(\epsilon_{ijk}{\Bbb L}^k -
                                \delta_{ij}{\Bbb L}^0), \cr
{\Bbb T}^i_2 . {\Bbb L}^0 &=& {i\over 2} {\Bbb L}^i.
\label{eq:G2}\eea
}
${\cal G}_1$ acts on ${\cal Q}_R$ like ${\cal G}_2$ does on ${\cal Q}_L$,
according to eq.~(\ref{eq:G2}); this clearly shows the difference between
${\cal G}_1$ and a ``left'' $SU(2)_L$ since by construction it also acts on
right-handed fermions.
${\cal G}_2$ acts on ${\cal Q}_R$ like ${\cal G}_1$ does on ${\cal Q}_L$,
according to eq.~(\ref{eq:G1}), and the same remark as above applies to it.

If one changes ${\cal Q}_L$ into $\overline{{\cal Q}_L}$, eqs.~(\ref{eq:G1}) and
(\ref{eq:G2}) are swapped; the same occurs with ${\cal Q}_R$.

The last properties just reflect the chiral structure of the symmetry under
consideration.

${\cal Q}_L$ is a reducible representation of each of these two groups and
can be decomposed into two spin $1/2$ doublets:\l
- two doublets of ${\cal G}_1$:
\begin{equation}
l_{1} = \left( \ba{c}  {1\over\sqrt{2}}({\Bbb L}^3 + i {\Bbb L}^0) \cr
                    {\Bbb L}^-
     \ea \right)
    = \left( \ba{c} \nu \cr
                    \ell^-
     \ea \right)_L,\qquad
l'_{1} = \left( \ba{c} {\Bbb L}^+ \cr
                   {1\over\sqrt{2}}({\Bbb L}^3 - i {\Bbb L}^0)
      \ea \right)
    = \left( \ba{c} \ell^+ \cr
                    \nu^c
       \ea \right)_L,
\label{eq:G1doublets}\end{equation}
with the group action

\vbox{
\bea
& &{\Bbb T}_1^3.\ell_L^- = - {1\over 2} \ell_L^-,\quad
{\Bbb T}_1^3.\ell_L^+ = {1\over 2} \ell_L^+,\quad
{\Bbb T}_1^3.\nu_L = {1\over 2}\nu_L,\quad
{\Bbb T}_1^3.(\nu^c)_L = -{1\over 2}(\nu^c)_L,\cr
& &{\Bbb T}_1^+.\ell_L^- = \nu_L,\quad
{\Bbb T}_1^+.\ell_L^+ = 0,\quad
{\Bbb T}_1^+.\nu_L = 0,\quad
{\Bbb T}_1^+.(\nu^c)_L = -\ell_L^+,\cr
& &{\Bbb T}_1^-.\ell_L^- = 0,\quad
{\Bbb T}_1^-.\ell_L^+ =-(\nu^c)_L,\quad
{\Bbb T}_1^-.\nu_L = \ell_L^-,\quad
{\Bbb T}_1^-.(\nu^c)_L = 0.
\label{eq:G1action}\eea
}

${\cal G}_1$ acts on $l_{1}$ like the $SU(2)_L$ group of the Standard
Model;

- two doublets of ${\cal G}_2$:
\begin{equation}
    l_{2} = \left( \ba{c} \nu^c \cr
                    \ell^-
     \ea \right)_L,\qquad
    l'_{2} =  \left( \ba{c} \ell^+ \cr
                    \nu
       \ea \right)_L,
\end{equation}
with the group action

\vbox{
\bea
& &{\Bbb T}_2^3.\ell_L^- = -{1\over 2} \ell_L^-,\quad
{\Bbb T}_2^3.\ell_L^+ = {1\over 2} \ell_L^+,\quad
{\Bbb T}_2^3.\nu_L = -{1\over 2}\nu_L,\quad
{\Bbb T}_2^3.(\nu^c)_L = {1\over 2}(\nu^c)_L,\cr
& &{\Bbb T}_2^+.\ell_L^- = (\nu^c)_L,\quad
{\Bbb T}_2^+.\ell_L^+ = 0,\quad
{\Bbb T}_2^+.\nu_L = -\ell_L^+,\quad
{\Bbb T}_2^+.(\nu^c)_L = 0,\cr
& &{\Bbb T}_2^-.\ell_L^- = 0,\quad
{\Bbb T}_2^-.\ell_L^+ =-\nu_L,\quad
{\Bbb T}_2^-.\nu_L = 0,\quad
{\Bbb T}_2^-.(\nu^c)_L = \ell_L^-;
\label{eq:G2action}\eea
}

With respect to the diagonal $SU(2)_V$ subgroup $\ti{\cal G}$ of the chiral
group ${\cal G}_1 \times {\cal G}_2$ with generators
$\ti{\Bbb T}^i = {\Bbb T}^i_1 + {\Bbb T}^i_2$, it decomposes into one spin
$1$ triplet, $\vec{\Bbb L}$, plus one singlet ${\Bbb L}^0$, with the group
action:\l

\vbox{
\bea
& &\ti{\Bbb T}^3.\ell_L^- = -\ell_L^-,\quad
\ti{\Bbb T}^3.\ell_L^+ = \ell_L^+,\quad
\ti{\Bbb T}^3.\nu_L = 0,\quad
\ti{\Bbb T}^3.(\nu^c)_L = 0,\cr
& &\ti{\Bbb T}^+.\ell_L^- = \nu_L + (\nu^c)_L,\quad
\ti{\Bbb T}^+.\ell_L^+ = 0,\quad
\ti{\Bbb T}^+.\nu_L = -\ell_L^+,\quad
\ti{\Bbb T}^+.(\nu^c)_L = -\ell_L^+,\cr
& &\ti{\Bbb T}^-.\ell_L^- = 0,\quad
\ti{\Bbb T}^-.\ell_L^+ = -(\nu_L + (\nu^c)_L),\quad
\ti{\Bbb T}^-.\nu_L = \ell_L^-,\quad
\ti{\Bbb T}^-.(\nu^c)_L = \ell_L^-.
\label{eq:GVaction}\eea
}

$\ti{\cal G}$ is the custodial symmetry which occurs in the mesonic sector.
The generator of the $U(1)$ group of electromagnetism is the ``$z$''
generator of this angular momentum.

When operating in the 4-dimensional vector space spanned by the four entries
of ${\cal Q}_L$, its three generators  write as $4\times 4$ matrices
according to (in the basis $({\Bbb L}^0, {\Bbb L}^3, {\Bbb L}^+ {\Bbb L}^-)$):
\begin{equation}
\tilde{\Bbb T}^+ = \left( \ba{cccc}
                   0  &  0        &  0        &  0  \cr
                   0  &  0        & -\sqrt{2} &  0  \cr
                   0  &  0        &  0        &  0  \cr
                   0  &  \sqrt{2} &  0        &  0
\ea\right),\quad
\tilde{\Bbb T}^- = \left( \ba{cccc}
                   0  &  0        &  0       &  0   \cr
                   0  &  0        &  0       &  \sqrt{2} \cr
                   0  & -\sqrt{2} &  0       &  0 \cr
                   0  &  0        &  0       &  0
\ea\right),\quad
\tilde{\Bbb T}^3 = \left( \ba{cccc}
                   0  &  0        &  0       &  0  \cr
                   0  &  0        &  0       &  0  \cr
                   0  &  0        &  1       &  0  \cr
                   0  &  0        &  0       & -1
\ea\right).
\label{eq:SU2Vlept}\end{equation}
The electric charge generator is identical with the third generator of
$SU(2)_V$:
\begin{equation}
{\Bbb Q} = \ti{\Bbb T}^3.
\end{equation}
The decompositions above apply to ${\cal Q}_R$ too; eqs.~(\ref{eq:G1action})
and (\ref{eq:G2action}) have to be swapped, but (\ref{eq:GVaction}) stays
unchanged.

We have thus achieved our first goal to define a ``chiral'' $SU(2)$
structure which acts on special representations of leptons in such a way
that its diagonal subgroup includes the electromagnetic $U(1)$.

\subsubsection{Invariants.}

They are constructed along the remark made in the previous paragraph that
changing ${\cal Q}_L$ into $\overline{{\cal Q}_L}$ swaps the role of
eqs.~(\ref{eq:G1}) and (\ref{eq:G2}), and that changing ${\cal Q}_L$ into
${\cal Q}_R$ has the same effect.

The unique quadratic expression invariant by ${\cal G}_1$ and ${\cal G}_2$ is
then
\begin{equation}
{\cal I} = \ol{\cal Q} {\cal Q}
     =  \ol{{\cal Q}_R} {\cal Q}_L + \ol{{\cal Q}_L} {\cal Q}_R .
\label{eq:invarlept}\end{equation}
It is of course also invariant by the diagonal $SU(2)_V$.

\subsubsection{A vector-like electroweak Lagrangian for leptons.}
 
Let us start, according to \cite{BellonMachet}, from the purely vectorial
Lagrangian

\vbox{
\bea
{\cal L} = & &i \ol{\ell^-} \gamma^\mu \p_\mu \ell^-
                  + i \bar \nu \gamma^\mu \p_\mu \nu\cr
           &+& {e\over \sqrt{2}s_W}\left(
           \ol{\ell^-} \gamma^\mu  W_\mu^- \nu
      +    \ol\nu \gamma^\mu  W_\mu^+ \ell^-
                             \right)\cr
             &-& {e\over 2s_W}\left(
                 \ol{\ell^-} \gamma^\mu W_\mu^3\ell^-
             - \ol\nu \gamma^\mu W_\mu^3\nu \right)\cr
             &-& {e\over 2c_W} \left( \ol{\ell^-} \gamma^\mu  B_\mu \ell^-
                               + \ol\nu \gamma^\mu  B_\mu \nu\right),
\label{eq:L}
\eea
}

to which we add the mass term
\begin{equation}
{\cal L}_m
=   -{m\over 2} \left( \ol{{\cal Q}_R} {\cal Q}_L + \ol{{\cal Q}_L} {\cal
Q}_R \right).
\label{eq:Lm}\end{equation}
The quadratic expression (\ref{eq:invarlept}) being invariant by both ${\cal
G}_1$ and ${\cal G}_2$, ${\cal L}_m$ is invariant by the chiral group ${\cal
G}_1 \times {\cal G}_2$, and
this invariance is independent of the mass $m$, which can vary with the
leptonic generation.

${\cal L}_m$ corresponds to a Dirac mass term. It is an important actor in
the
``see-saw'' mechanism evoked in the last section. A Majorana mass term for
the
neutrino would correspond to the combination (forgetting the charged
leptons) $(\ol{{\Bbb R}^3} {\Bbb L}^3 + \ol{{\Bbb L}^3} {\Bbb R}^3)
- (\ol{{\Bbb R}^0} {\Bbb L}^0 + \ol{{\Bbb L}^0} {\Bbb R}^0) + \cdots$,
(the ``$-$'' sign makes the difference), which is not invariant by ${\cal
G}_1$.

Using the properties of charge conjugation, it turns out \cite{Machet3}
that $\cal L$ is
the sum ${\cal L} = {\cal L}_1 + {\cal L}'_1$ (we use the abbreviated
notation $\gamma_{\mu L} = \gamma_\mu (1-\gamma_5)/2, \gamma_{\mu R} =
\gamma_\mu(1+\gamma_5)/2$)

\vbox{
\bea
{\cal L}_1 = & &i \ol{\ell^-} \gamma^\mu_L \p_\mu \ell^-
                  + i \bar \nu \gamma^\mu_L \p_\mu \nu\cr
             &+& {e\over \sqrt{2}s_W}\left(
           \ol{\ell^-} \gamma^\mu_L  W_\mu^- \nu
           + \ol\nu \gamma^\mu_L  W_\mu^+ \ell^-
                             \right)\cr
             &-& {e\over 2s_W}\left(
                 \ol{\ell^-} \gamma^\mu_L W_\mu^3\ell^-
             - \ol\nu \gamma^\mu_L W_\mu^3\nu \right)\cr
             &-& {e\over 2c_W} \left( \ol{\ell^-} \gamma^\mu_L  B_\mu \ell^-
                               + \ol\nu \gamma^\mu_L B_\mu \nu \right);
\label{eq:L1}
\eea
}

\vbox{
\bea
{\cal L}'_1 = & &i \ol{\ell^+} \gamma^\mu_L \p_\mu \ell^+
                  + i \ol{\nu^c} \gamma^\mu_L \p_\mu \nu^c\cr
             &-& {e\over \sqrt{2}s_W}\left(
            \ol{\ell^+} \gamma^\mu_L  W_\mu^+ \nu^c
       -    \ol{\nu^c} \gamma^\mu_L  W_\mu^- \ell^+
                             \right)\cr
             &+& {e\over 2s_W}\left(
                 \ol{\ell^+} \gamma^\mu_L W_\mu^3\ell^+
             - \ol{\nu^c} \gamma^\mu_L W_\mu^3\nu^c \right)\cr
             &+& {e\over 2c_W} \left( \ol{\ell^+} \gamma^\mu_L B_\mu \ell^+
                          + \ol{\nu^c} \gamma^\mu_L B_\mu \nu^c \right).
\label{eq:L'1}
\eea
}

corresponding respectively to the two doublets $\ell_1$ and $\ell'_1$, and
that it can also be rewritten
\begin{equation}
{\cal L} = i\overline{{\cal Q}_L} \gamma_\mu D^\mu_{{\cal G}_1 \times
U(1)}{\cal Q}_L,
\label{eq:Lagr1}
\end{equation}
where $D^\mu_{{\cal G}_1 \times U(1)}$ is the covariant derivative with
respect to ${\cal G}_1 \times U(1)$; the $U(1)$ is defined by a
generator $\Bbb Y$ satisfying the Gell-Mann-Nishijima relation
\begin{equation}
{\Bbb Y} = {\Bbb Q} - {\Bbb T}^3_1;
\label{eq:GMN1}
\end{equation}
and corresponds to a gauging of the leptonic number; indeed, the leptonic
numbers of the entries of $\cal Q$ are $(-2) \times $ their $U(1)$ quantum
numbers.

The same result can be obtained by considering ${\cal Q}_R$ instead of
${\cal Q}_L$.

Would we make a similar construction with the group ${\cal G}_2 \times
U(1)$, we would obtain a Lagrangian similar to (\ref{eq:L}) but with $\nu$
and $\nu^c$ swapped, which leaves the kinetic terms unaltered.

\subsubsection{Implementing the custodial \boldmath{$SU(2)_V$} symmetry.}

While the mass term is trivially ${\cal G}_1 \times {\cal G}_2$
chirally invariant, this is not the case for
the  kinetic term (\ref{eq:Lagr1}) for ${\cal Q}_L$.

Indeed, as already mentioned, ${\cal Q}_L$ transforms by ${\cal G}_1$
according to eq.~(\ref{eq:G1}) and by ${\cal G}_2$ according to
eq.~(\ref{eq:G2}), but $\overline{{\cal Q}_L}$ transforms with
eqs.~(\ref{eq:G1}) and (\ref{eq:G2}) swapped. So, the kinetic
term (\ref{eq:Lagr1}) is neither ${\cal G}_1$ nor ${\cal G}_2$-invariant.

This difference with respect to the mesonic case explains why now the
custodial $SU(2)_V$ symmetry in particular is not something which is
automatically achieved when $g' \rightarrow 0$, but requires some
constraint to be satisfied.

We can rewrite ${\cal L}_1 + {\cal L}'_1$ in the form

\vbox{
\bea
& &2({\cal L}_1 + {\cal L}'_1) =  \cr
 & &i(\ol{{\Bbb L}^+} \gamma_\mu \p^\mu {\Bbb L}^+
                +  \ol{{\Bbb L}^-} \gamma_\mu \p^\mu {\Bbb L}^-
                +  \ol{{\Bbb L}^3} \gamma_\mu \p^\mu {\Bbb L}^3
                +  \ol{{\Bbb L}^0} \gamma_\mu \p^\mu {\Bbb L}^0)\cr
&+&{g} \left(
\ol{{\Bbb L}^-} \gamma_{\mu L} ({1\over \sqrt{2}}
                      (W_\mu^- \tilde{\Bbb T}^+ +W_\mu^+ \tilde{\Bbb T}^-)
                            +{Z_\mu \over c_W} \tilde{\Bbb T}^3).{\Bbb L}^-
\right.\cr
& & \left. \hphantom{aaaaaaaaaaaaaaaaaaaaaaaaaaaaa}
+\ol{{\Bbb L}^+} \gamma_{\mu L} ({1\over \sqrt{2}}
                      (W_\mu^- \tilde{\Bbb T}^+ +W_\mu^+ \tilde{\Bbb T}^-)
                            +{Z_\mu \over c_W} \tilde{\Bbb T}^3).{\Bbb L}^+
\right.\cr
& & \left. \hphantom{aaaaaaaaaaaaaaaaaaaaaaaaaaaaa}
 +\ol{{\Bbb L}^3} \gamma_{\mu L} ({1\over \sqrt{2}}
   (W_\mu^- \tilde{\Bbb T}^+ + W_\mu^+ \tilde{\Bbb T}^-)
     +{Z_\mu \over c_W} \tilde{\Bbb T}^3). {\Bbb L}^3
               \right)\cr
&+& i\;{g} \left(
  - \ol{{\Bbb L}^0} (\gamma_{\mu L} {Z_\mu \over c_W} {\Bbb L}^3
                    + \gamma_{\mu L} W_\mu^- {\Bbb L}^+
                    + \gamma_{\mu L} W_\mu^+ {\Bbb L}^-) \right. \cr
& & \left. \hphantom{aaaaaaaaaaaaaaaaaaaaaaaaaaaaa}
           + (\ol{{\Bbb L}^3} \gamma_{\mu L} {Z_\mu \over c_W}
        +\ol{{\Bbb L}^-} \gamma_{\mu L} W_\mu^-
        + \ol{{\Bbb L}^+} \gamma_{\mu L} W_\mu^+){\Bbb L}^0
               \right)\cr
&+& {g'} (\ol{{\Bbb L}^+} \gamma_{\mu L}  B^\mu {\Bbb L}^+
           -\ol{{\Bbb L}^-} \gamma_{\mu L}  B^\mu {\Bbb L}^-),
\label{eq:Lag}\eea
}

where we have used the fact that $\tilde{\Bbb T}^+$ does not act on ${\Bbb
L}^+$, nor $\tilde{\Bbb T}^-$ on ${\Bbb L}^-$, nor $\tilde{\Bbb T}^3$ on
${\Bbb L}^3$.

The pure kinetic terms and the second line of (\ref{eq:Lag})
are  globally $SU(2)_V$ invariant when the triplet of gauge bosons
$W_\mu^\pm$ and $Z_\mu/c_W$ transform (see subsection
\ref{subsec:custodial})
like a vector in the adjoint representation of this group.

The next line of (\ref{eq:Lag}) is also globally $SU(2)_V$ invariant,
since ${\Bbb L}^0$ and $\ol{{\Bbb L}^0}$ are singlets and are each
multiplied by another singlet made by the scalar product of two triplets.

Now, $B_\mu$ being considered (see subsection \ref{subsec:custodial})
as a singlet of $SU(2)_V$,
the last line of (\ref{eq:Lag}) only becomes $SU(2)_V$
invariant if, as can be seen by performing an
explicit transformation and using (\ref{eq:GVaction},\ref{eq:SU2Vlept})
\begin{equation}
\nu  +\nu^c = 0,
\label{eq:Majorana}
\end{equation}
{\em i.e.} the neutrino has to be a Majorana particle,
with only one helicity (or
chirality), which can be written \cite{Ramond}, in the 4-component notation,
either $\gamma_5\chi$ or  $\gamma_5\omega$ with
\begin{equation}
\chi = \left( \ba{c} \psi_L \cr
                      -\sigma^2 \psi_L^\ast \ea \right),\quad
\omega = \left( \ba{c} \psi_R \cr
                      -\sigma^2 \psi_R^\ast \ea \right).
\label{eq:chiomega}\end{equation}
$\psi_L$ (resp. $\psi_R$)  is a two-component Weyl spinor transforming like
a $(1/2,0)$ (resp. $(0,1/2)$) representation of the Lorentz group;
$\sigma^2$ is the second Pauli matrix and the superscript ``$\ast$'' means
``complex conjugation''; $\sigma^2 \psi_L^\ast$ (resp. $\sigma^2
\psi_R^\ast$) transforms like a $(0,1/2)$ (resp. $(1/2,0)$) representation.

We thus conclude that:\l
{\em The leptonic Lagrangian ${\cal L}$ can have a global custodial
$SU(2)_V$ symmetry only if  the neutrino is a Majorana particle.}

Clearly, this condition is not compatible with the decomposition
(\ref{eq:G1doublets}) and the corresponding laws of transformation
(\ref{eq:G1}). In
particular, it requires that the $U(1)$ leptonic number be not conserved.
We shall see in the next subsection how the necessary modifications can occur
dynamically with the introduction of a ``hidden'' sector, along the lines of
\cite{BellonMachet}.

\subsection{From a vectorial theory to an effective $\mathbf{V-A}$ theory
with a decoupled right-handed neutrino.}
\label{subsec:V-A}

The goal of going from a fundamental vectorial electroweak theory of leptons
to an effective $V-A$ interaction as we observe it can be achieved
\cite{BellonMachet} by
introducing a scalar triplet of composite scalars, ``made of'' leptons,
and the neutral component of which gets a non-vanishing vacuum expectation
value. These composite scalars not being independent degrees of freedom,
one must introduce, in the quantization process, constraints in the Feynman
path integral, which can be exponentiated
into an effective Lagrangian. It can be treated at leading order in an
expansion in powers of $1/N$ \cite{BellonMachet} and introduces a drastic
asymmetry between the two (Majorana) neutrinos: it gives one of them an
infinite mass, and this one is consequently unobservable, while the
other, by an exact see-saw  mechanism, gets a
vanishing mass.

The effective Lagrangian of constraint includes four-leptons couplings, but
their effective value go to zero in the limit of decoupling
neutrino, preserving renormalizability at the approximation that we are
working at.

The infinitely massive neutrino conspires with the vanishing
effective four-fermion coupling to alter, at the one-loop level, the bare
leptonic couplings to reconstruct the well known $V-A$ effective structure
of weak currents.

Furthermore, the composite scalar finally decouple, playing the role of a
hidden sector. 

So, the two mysterious phenomena of the non-observation of a right-handed
neutrino and of the $V-A$ structure of weak currents (parity violation) have
been given the same origin to yield,  at the approximation of leading order
in $1/N$, an effective interaction indistinguishable from those arising from
the Glashow-Salam-Weinberg model.

We will only here sketch out the main steps of the demonstration.

\subsubsection{Introducing a composite scalar triplet.}

We rewrite the Lagrangian (\ref{eq:L}) plus the mass term (\ref{eq:Lm})
in terms of the Majorana neutrinos $\chi$ and $\omega$ conveniently
reexpressed as
\bea
\chi &=& \nu_L + (\nu_L)^c,\cr
\omega &=& \nu_R + (\nu_R)^c;
\eea
this yields

\vbox{
\bea
{\cal L} + {\cal L}_m &=&
 i \ol{\ell^-} \gamma^\mu \p_\mu \ell^-
+ {i\over 2} \bar\chi \gamma^\mu \p_\mu \chi
+ {i\over 2} \bar\omega \gamma^\mu \p_\mu \omega\cr
&+& {e\over\sqrt{2} s_W}(\ol{\ell^-} \gamma^\mu_L W_\mu^- \chi
                   +\bar\chi \gamma^\mu_L W_\mu^+ \ell^-)\cr
&+& {e\over\sqrt{2} s_W}(\ol{\ell^-} \gamma^\mu_R W_\mu^- \omega
                   +\bar\omega \gamma^\mu_R W_\mu^+ \ell^-)\cr
&+& {e\over 2 s_W}\left(
\bar\chi \gamma^\mu_L W_\mu^3\chi
+  \bar\omega  \gamma^\mu_R  W_\mu^3\omega
  -\ol{\ell^-} \gamma^\mu W_\mu^3 \ell^- \right)\cr
&-& {e\over 2c_W} \left(\ol{\ell^-} \gamma^\mu B_\mu \ell^-
    + \bar\chi \gamma^\mu_L B_\mu \chi
    + \bar\omega \gamma^\mu_R B_\mu \omega \right)\cr
&-& {m\over 2}(\bar\chi \omega + \bar\omega \chi + 2 \ol{\ell^-} \ell^-).
\label{eq:Lmajo}\eea
}

({\em Remark:} would we have built the model with the group ${\cal G}_2
\times U(1)$, we would have obtained, instead of eq.~(\ref{eq:Lmajo})
$\hat{\cal L}$, deduced from $\cal L$ by the exchange of $\chi$ and
$\omega$, or, equivalently, by that of the ``left'' and ``right''
projectors.)

We introduce a scalar composite triplet $\Delta$ with leptonic number $2$:
\begin{equation}
\Delta = \left( \ba {l} \Delta^{0} \cr
                        \Delta^{-} \cr
                        \Delta^{--} \cr
         \ea \right)
       ={\rho\over\nu^3}
          \left( \ba {c} \ol{\omega_L} \omega_R \cr
                          {1\over \sqrt{2}}(\ol{\ell^+_L} \omega_R+
                                    \ol{\omega_L} \ell^-_R)\cr
                          \ol{\ell^+_L} \ell^-_R
          \ea \right)
       = {\rho\over\nu^3}
          \left( \ba {c} \ol{\nu^c}\ {1+\gamma_5\over 2}\ \nu \cr
                    {1\over \sqrt{2}}(\ol{\ell^+}\ {1+\gamma_5\over 2}\ \nu
+
                               \ol{\nu^c}\ {1+\gamma_5\over 2}\ \ell^-) \cr
                          \ol{\ell^+}\ {1+\gamma_5\over 2}\ \ell^-
          \ea \right).
\end{equation}
It is a triplet of ${\cal G}_1$ but not a representation of ${\cal G}_2$,
nor of $\ti{\cal G}$.

Its hermitian conjugate is:
\begin{equation}
\ol\Delta = \left( \ba {l} {\ol{\Delta^0}} \cr
                            \Delta^{+} \cr
                            \Delta^{++} \cr
         \ea \right)
       ={\rho\over\nu^3}
          \left( \ba {c} \ol{\omega_R} \omega_L \cr
                          {1\over \sqrt{2}}(\ol{\ell^-_R} \omega_L +
                                   \ol{\omega_R} \ell^+_L) \cr
                          \ol{\ell^-_R} \ell^+_L
          \ea \right)
       = {\rho\over\nu^3}
          \left( \ba {c} \ol{\nu}\ {1-\gamma_5\over 2}\ \nu^c \cr
                 {1\over \sqrt{2}}(\ol{\ell^-}\ {1-\gamma_5\over 2}\ \nu^c +
                               \ol{\nu}\ {1-\gamma_5\over 2}\ \ell^+) \cr
                          \ol{\ell^-}\ {1-\gamma_5\over 2}\ \ell^+
          \ea \right).
\end{equation}
As soon as the mass of $\omega$ is non vanishing, electroweak vacuum
fluctuations like described in fig.~3

\vbox{
\begin{center}
\epsfig{file=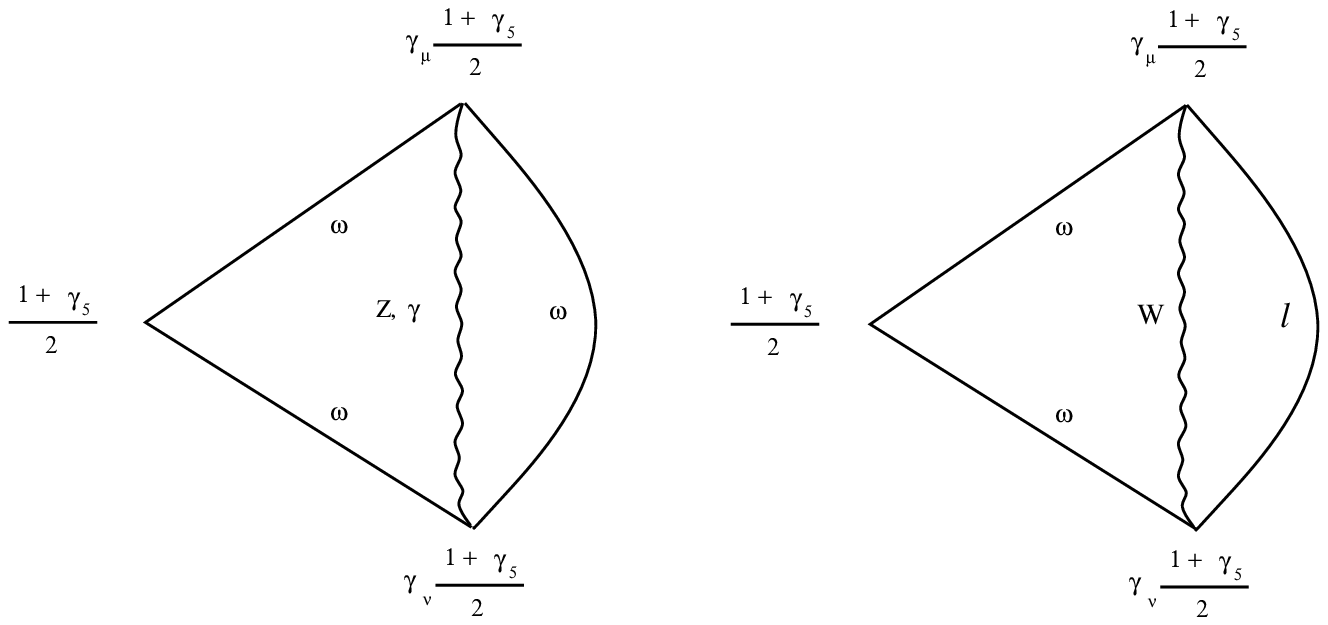}
\end{center}
\centerline{{\em Fig.~3: contributions to the vacuum expectation value of
$\Delta^0$.}}
\figskip
}

can trigger
\begin{equation}
\la\Delta^0\ra = \la\ol{\Delta^0}\ra = \rho.
\end{equation}
The choice of an $\la\bar\omega \omega\ra$ condensate, breaking the symmetry
between $\chi$ and $\omega$, spontaneously breaks
the ``left-right'' symmetry, or, equivalently, parity.

It could be thought arbitrary since the same type of vacuum
fluctuations can also {\em a priori} trigger $\la \bar\chi \chi \ra \not=
0$. However, the diagrams under consideration vanish with the mass of the
internal fermion. As, by the see-saw mechanism evoked below, an
$\la \bar\omega \omega \ra$ condensate pushes the $\chi$ mass to $0$ at
the same time that is pushes the $\omega$ mass to $\infty$,
the $\la \bar\chi \chi \ra$ condensate is then automatically suppressed, and
vice-versa. This qualitative explanation forbids the coexistence of both
condensates.

The proposed mechanism can also be interpreted along the following lines:
by expanding
$\gamma_\mu$ into $(\gamma_{\mu L} + \gamma_{\mu R})$, the vectorial
Lagrangian (\ref{eq:L}) can be considered to be that of an
$SU(2)_L \times SU(2)_R \times U(1)$ gauge model for the doublet
$(\nu, \ell^-)$; both $SU(2)$'s
act the same way, with the ``left''and ``right'' gauge fields identified.
The ``Higgs'' multiplet $\Delta$ being a  triplet of $SU(2)_R$
and of $SU(2)_L$ with a non-vanishing leptonic number,  the
condensation of its neutral component spontaneously breaks both $SU(2)$'s,
and  $U(1)$.

\subsubsection{The Lagrangian of constraint.}

To take into account the non-independence of the leptonic and
$\Delta$ degrees of freedom, we introduce constraints which, once
exponentiated, yield the effective Lagrangian ($\Lambda$ is an arbitrary
mass scale):
\begin{equation}\ba{lcl}
{\cal L}_c &=& \lim_{\beta\rar 0}
            -{\Lambda^2\over 2\beta}\left[
\left( \Delta^0 -{\rho\over\nu^3}\, \ol{\omega_L} \omega_R \right)
\left(\ol{\Delta^0}-{\rho\over\nu^3}\, \ol{\omega_R} \omega_L\right)
\right.\cr
& &+
\left.\left(\Delta^- -{1\over\sqrt{2}}{\rho\over\nu^3}(\ol{\omega_L}
\ell^-_R +
\ol{\ell^+_L} \omega_R) \right)
\left(\Delta^+ -{1\over\sqrt{2}}{\rho\over\nu^3}(\ol{\ell^-_R} \omega_L +
\ol{\omega_R} \ell^+_L) \right)\right.\cr
& &+
\left.\left (\Delta^{--} -{\rho\over\nu^3}\, \ol{\ell^+_L} \ell^-_R\right)
\left(\Delta^{++} -{\rho\over\nu^3}\, \ol{\ell^-_R} \ell^+_L\right)
\right],
\label{eq:Lc}\ea\end{equation}
where $\Lambda$ is an arbitrary mass scale.

\subsubsection{Effective 4-leptons couplings and mass eigenstates.}

The equations (\ref{eq:L}) and (\ref{eq:Lc}) yield a ``see-saw'' mechanism
\cite{GellMannRamondSlansky} in the neutrino sector. Indeed, when
$\la\Delta^0\ra = \rho$, ${\cal L}_c$ gives
the $\omega$ neutrino an infinite bare Majorana mass,
\begin{equation}
M_0 = -\frac {\Lambda^2 \rho^2}{\beta\nu^3},
\end{equation}
the $\chi$ neutrino a vanishing (Majorana) mass, and the finite Dirac mass
of the mass Lagrangian (\ref{eq:Lm}) connects $\chi$ and $\omega$.
We have to diagonalize the mass matrix to get the mass eigenstates,
(see for example \cite{ChengLi}); they are the Majorana neutrinos
$\chi$ and $\omega$ themselves, and correspond to mass eigenvalues
$0$ and $\infty$ respectively.  The charged lepton keeps its Dirac mass $m$.

However, 4-fermions couplings may alter the mass spectrum, together with
being an obstacle for renormalizability. We propose to build a reshuffled
perturbative expansion based not on the `bare' (infinite when $\beta
\rightarrow 0$) 4-fermions couplings occurring
in ${\cal L}_c$, but rather on effective couplings obtained by resumming
infinite series of `ladder' diagrams as proposed by Nambu and Jona-Lasinio
\cite{NambuJonaLasinio}. This corresponds to only keeping the leading order
in an expansion in powers of $1/N$. We however differ from them by
bare couplings and a bare fermion mass both infinite; this
makes the effective 4-fermions couplings vanish with $\beta$ like $\beta^2$,
and the ``see-saw'' mechanism above stay unaltered.

\subsubsection{The scalars and their decoupling.}

$SU(2)_L$ is broken by the condensation of $\Delta^0$, which
thus weakly contributes to the masses of the gauge fields.
If $v/\sqrt{2}$ if the vacuum expectation value of the hadronic Higgs boson
we impose its role to be dominant, which yields the necessary condition
\begin{equation}
                          \rho <\!\! < v
\end{equation}
consistent with an electroweak nature for $\rho$ (see fig.~3), while that of
$v$ lies {\sl a priori} outside the realm of these interactions.

One then has to shifting the usual way the neutral scalar field according to
\begin{equation}
                          \Delta^0 = \rho + \delta^0.
\end{equation}
{}From the expression of ${\cal L}_c$, we see that the non-vanishing of $\rho$
yields an infinite mass for $\delta^0$, $\Delta^+$, $\Delta^{++}$ and their
conjugates.

None of the components of $\Delta$ appears as an asymptotic state
and we do not require electric charge quantization for them.
It is the same kind of implicit assumption that we made in the
mesonic sector where explaining charge quantization for quarks and their
underlying gauge theory (Quantum Chromodynamics) was not sought for.
So, we can allow a Lagrangian which is not globally $\ti{\cal G}$ invariant
in the hidden sector; this can be the case for the kinetic term for
$\Delta$; this non-invariance is responsible for that of the
Lagrangian of the Glashow-Salam-Weinberg model.

By the decoupling of the weak hidden sector, the Goldstones of the broken
$SU(2)_L$ symmetry align with the customary hadronic ones.
This decoupling also motivates non introducing
other triplets of composite states, with $\omega$ replaced by $\chi$, since
they would not modify the result: as soon as only one type of
condensate can occur, the non-condensing additional scalars would simply
fade away without any visible effect.

\subsubsection{The effective $V-A$ theory.}

The (massless) $\chi$ neutrino has the standard weak $V-A$ couplings and we
can identify it with the observed neutrino.

The $\omega$ neutrino is infinitely massive and will never be produced as
asymptotic state; we however expect renormalization effects
through $\omega$ loops \cite {ApplequistCarazzone}.
They drastically affect the neutral weak couplings of the
leptons, in a way that rebuilds their ``standard'' $V-A$ structure. This result,
non-trivial if one remembers that the original coupling is purely vectorial,
is sketched out below (see \cite{BellonMachet}).

To the bare (purely vectorial) couplings, we must add the following diagrams
(in figures below, the $L$, $R$'s at the vertices stand for the
projectors $(1-\gamma_5)/2$ and $(1+\gamma_5)/2$)

\vbox{
\figskip
\begin{center}
\epsfig{file=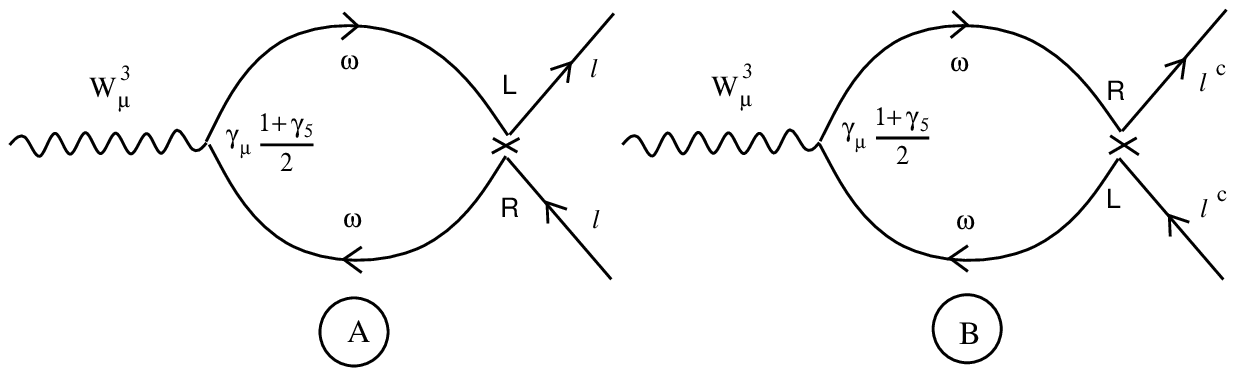}
\end{center}
\centerline{{\em Fig.~4: additional leptonic neutral couplings.}}
\figskip
}

In fig.~4, the 4-fermions vertex is the effective $(\ell\ell\omega\omega)$
coupling vanishing like $\beta^2$.
Because of this dependence in $\beta$, diagrams similar to fig.~4 but
with $\omega$ replaced with $\chi$ vanish. Those involving
$\omega$ do not because the $\omega$ loop behaves like $\beta^{-2}$,
and yield  a coupling
\begin{equation}
- W_\mu^3\, \ol\ell\,\gamma^\mu {1+\gamma_5\over 2}\ell,
\end{equation}
such that  $W_\mu^3$ couples finally to
\begin{equation}
\ol\ell\,\gamma^\mu \ell -\ol\ell\,\gamma^\mu {1+\gamma_5\over 2}\, \ell
                      = \ol\ell\,\gamma^\mu {1-\gamma_5\over 2}\, \ell,
\end{equation}
which is the ``standard'' $V-A$ coupling.

The charged couplings do not get modified with respect to their bare values
since the diagram equivalent to fig.~4 depicted in fig.~5 involving an
infinitely massive
$\omega$ behaves like $\beta^2 M_0\ln M_0$ and so vanishes with $\beta$.

\vbox{
\figskip
\begin{center}
\epsfig{file=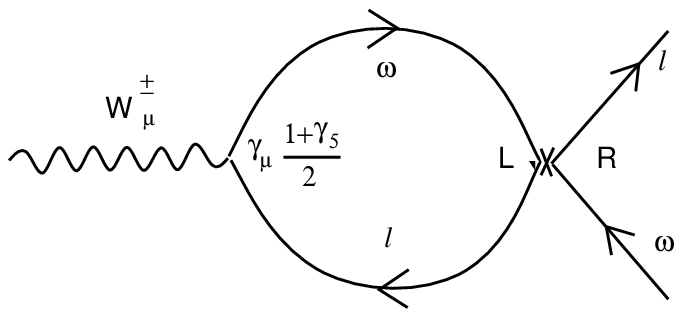}
\end{center}
\centerline{\em Fig.~5: the charged couplings do not get altered.}
\figskip
}

At the approximation that we are working at, the leading order in $1/N$,
ours is thus presently experimentally indistinguishable from the Standard
Model.

\section{Conclusion.}
\label{section:conclusion}

Symmetries have been our main concern in the proposal that we
made above for an extension (to mesons) and a modification (for leptons) 
of the standard electroweak model.

The $SU(2)_L \times SU(2)_R$ chiral symmetry and its breaking down to the
custodial diagonal $SU(2)_V$ has been seen to play a crucial
role in the mesonic sector and in its spectrum;
the new picture that arises alters in particular the usual framework in
which the pseudoscalar mesons are the Goldstones of the chiral 
$U(N)_L \times U(N)_R$ symmetry spontaneously broken down to the diagonal
$U(N)$ of flavour; it furthermore opens the
door, via the existence of soliton-like classical solutions, to the
existence of a strongly interacting sector which needs to be
investigated. 

We propose that, in relation with the extreme accuracy with which the
electric charge is observed to be quantized, that the custodial symmetry is
an exact global symmetry of the physics of mesons.

We have also seen how crucial is the determination of the $\epsilon '$
parameter in kaon physics to determine whether $CP$ is violated or not.
Our conclusion is indeed that the usual mechanism proposed at the fermionic
level to trigger $CP$ violation may not be operative anymore, which
reinforces the mystery of the origin of this phenomenon and leaves the door
open for other mechanisms \cite{LeeWeinberg}. If $\epsilon '$ if
found compatible with zero, we could conclude that $CP$ is not violated, and
that the observation that some electroweak mass eigenstates are not $CP$
eigenstates is just a reflection of parity violation.

The quarks, considered here as only mathematical, bear no more connection 
with leptons. They have always been anyhow totally
different types of objects, the former being only fields and not particles.

That the leptonic sector has the same custodial symmetry as the mesonic
sector seems mandatory if one wants that the electric charge of leptons is
quantized for the same reasons. This is proposed as the unifying link between
the two sectors, and it now holds between fields that are also
asymptotic states.

To consider the fundamental leptonic electroweak Lagrangian as purely
vectorial is not a new idea. It solves many conceptual problems, including
the one of anomalies. We could connect above  the observed parity violation
to another experimental mystery, the absence of a right-handed neutrino and
the vanishing (?) mass of the left-handed one \cite{Petcov} and give them a
common origin.  Of course a justification
remains to be found for artificially introducing, as we did, an
additional triplet of composite scalars ``made of'' leptons, and thus to
find the true origin of parity violation as we observe it.

The paraphernalia of field theory  are also to be used to make
phenomenological predictions about electroweak processes involving mesons.
As the model has been built to be compatible with the Glashow-Salam-Weinberg
for quarks, deviations, if any, should be rather subtle. But the
ideas of factorization could for example be tested precisely, and, if one
adds a simple model for strong interactions, results concerning $K
\rightarrow 2\pi$ decays can be expected. The latter are a particular good
test field  since the problem of the $\Delta I = 1/2$ rule is
precisely that of a strong breaking of the isospin symmetry in the final
state, which can be easily triggered in our framework.
The problems lie more in determining what is the nature of the observed
particles, strong or electroweak eigenstates, and which ones occur in the
internal lines of the corresponding diagrams. As the latter also involve the
Higgs field, it is not excluded that one gets some information about it
in this way, though, as the custodial symmetry is  more ``protective'' than
ever, any kind of decoupling theorem may well be exact now.

We hope that the reader has found here some new ideas in the pressing
hunt for a fundamental theory of the interactions of particles.

\vbox{
\smallskip{\em \underline{Acknowledgments}: it is a pleasure to thank the
organizing committee of the $2^{nd}$ International Symposium on Symmetries
in Subatomic Physics, and especially Prof. Ernest M. Henley, for their very
kind hospitality in Seattle and all their successful efforts to make this
meeting most interesting and enjoyable.}
}

\vfill\eject
\null
{\Large \bf Figure captions.}
\vskip 2cm
\begin{em}
{\obeylines
Fig.~1: The leptonic decay of a pseudoscalar meson;
Fig.~2: The semi-leptonic decay of a meson;
Fig.~3: contributions to the vacuum expectation value of $\Delta^0$;
Fig.~4: additional leptonic neutral couplings;
Fig.~5: the charged couplings do not get altered.
}
\end{em}
\vfill\eject

\begin{em}

\end{em}


\begin{thebibliography}{50}
%
\bibitem{GlashowSalamWeinberg}
       S. L. GLASHOW: Nucl. Phys. 22 (1961) 579;\l
       A. SALAM: in ``Elementary Particle Theory: Relativistic Groups and
             Analyticity'' (Nobel symposium No 8), edited by N. Svartholm
             (Almquist and Wiksell, Stockholm 1968);\l
       S. WEINBERG: ``A model of leptons'', Phys. Rev. Lett. 19 (1967) 1264.

\bibitem{Machet1}
       B. MACHET: ``Chiral Scalar Fields, Custodial Symmetry in
                   Electroweak $SU(2)_L \times U(1)$, and the Quantization
                   of the Electric Charge'',
                   hep-ph/9606239,  Phys. Lett. B 385 (1996) 198-208.

\bibitem{Olive}
         See for example:\l
        D. I. OLIVE: ``Exact electromagnetic duality'', invited talk at the
                      Trieste Conference on Recent Developments in Statistical
                      Mechanics and Quantum Field Theory (April 1995),
                      preprint SWAT/94-95/81 (1995), and references therein.

\bibitem{Machet2}
        B. MACHET: ``Two results concerning $CP$ violation for $J=0$ mesons'',
                     preprint PAR-LPTHE 97/18, hep-ph/9706308.

\bibitem{CabibboKobayashiMaskawa}
        N. CABIBBO: ``Unitary symmetry and leptonic decays'',
                           Phys. Lett. 10 (1963) 513;\l
        M. KOBAYASHI and T. MASKAWA:  ``$CP$-Violation in the Renormalizable
           Theory of Weak Interactions'', Prog. Theor. Phys. 49 (1973) 652.

\bibitem{CurrentAlgebra}
        See for example:\l
        S. L. ADLER and R. F. DASHEN: ``Current Algebra and Application
                              to Particle Physics'', (Benjamin, 1968);\l
        B. W. LEE: ``Chiral Dynamics'', (Gordon Breach, 1972),
                         and references therein.

\bibitem{Machet3}
        B. MACHET: ``Leptonic custodial symmetry, quantization of the
                    electric charge and the neutrino in the Standard Model'',
                    hep-ph/9606308, Mod. Phys. Lett. A 11 (1996) 2297-2307.

\bibitem{BouchiatIliopoulosMeyer}
        C. BOUCHIAT, J. ILIOPOULOS and Ph. MEYER: ``An anomaly-free version
              of Weinberg's model'', Phys. Lett. 38 B (1972) 519;\l
        D. J. GROSS and R. JACKIW: ``Effect of Anomalies on
                   Quasi-Renormalizable Theories'', Phys. Rev. D 6 (1972) 477.

\bibitem{BellonMachet}
        M. BELLON \& B. MACHET: ``The Standard Model of leptons as a purely
                    vectorial theory'', hep-ph/9305212,
                                       Phys. Lett. B 313 (1993) 141.
        
\bibitem{GellMann}
        M. GELL-MANN: ``A schematic model of baryons and mesons'',
                       Phys. Lett. 8 (1964) 214.

\bibitem{GellMannLevy}
        M. GELL-MANN \& M. LEVY: Nuov. Cim. 16 (1960) 705.

\bibitem{GellMannNishijima}
        M. GELL-MANN: Phys. Rev. 92 (1953) 833;\l
        K. NISHIJIMA: Prog. Theor. Phys. 13 (1955) 285.

\bibitem{GildenerWeinberg}
        E. GILDENER and S. WEINBERG: ``Symmetry breaking and scalar bosons'',
                                     Phys. Rev. D 13 (1976) 3333;\l
        E. GILDENER: ``Gauge-symmetry hierarchies'', Phys. Rev. D 14 (1976)
                                      1667.
\bibitem{ChoMaison}
        Y. M. CHO \& D. MAISON: `` Monopole configuration in Weinberg-Salam
                       model'', Phys. Lett. B 391 (1997) 360-365.

\bibitem{ChoKimm}
        Y. M. CHO \& K. KIMM: ``Electroweak Monopoles'', hep-th/9705213;\l
        Y. M. CHO \& K. KIMM: ``Finite Energy Electroweak Monopoles'',
                               hep-th/9707038.
\bibitem{Veltman}
        M. VELTMAN: Acta Phys. Pol. B8 (1977) 475;\l
        F. ANTONELLI, M. CONSOLI and O. PELLEGRINO: Nucl. Phys. B 183 (1981)
                                                    195;\l
        J. VAN DER BIJ and M. VELTMAN: Nucl. Phys. B 232 (1984) 205;\l
        J. VAN DER BIJ: Nucl. Phys. B 248 (1984) 141;\l
        M. B. EINHORN and J. WUDKA: Phys. Rev. D 39 (1989) 2758.

\bibitem{Sikivie}
        P. SIKIVIE, L. SUSSKIND, M. VOLOSHIN and V. ZAKHAROV: ``Isospin
                  breaking in technicolour models'', Nucl. Phys. B 173
                                                            (1980) 189.

\bibitem{Skyrme}
        T. H. R. SKYRME: Proc. Roy. Soc. A260 (1961) 127;\l
        E. WITTEN: ``Global aspects of Current Algebra'', Nucl. Phys. B 223
                 (1983) 422; ``Current Algebra, baryons, and quark
                  confinement'', ibidem 433;\l
        G. S. ATKINS, C. R. NAPPI and E. WITTEN: ``Static properties of the
               nucleon in the Skyrme model'',  Nucl. Phys. B 228 (1983) 552.

\bibitem{LeeQuiggThacker}
        B. W. LEE, C. QUIGG \& H. B. THACKER: `` Weak interactions at very
                         high energies: The role of the Higgs-boson mass'',
                         Phys. Rev. D 16 (1977) 1519.
       
\bibitem{ChristensonCroninFitchTurlay}
       J. H. CHRISTENSON, J. W. CRONIN, J. W. FITCH and R. TURLAY:
           ``Evidence for the $2\pi$ decay of the $K^0_2$ meson'',
             Phys. Rev. Lett. 13 (1964) 138;\l
       V. L. FITCH: ``The discovery of charge-conjugation parity asymmetry'',
             Rev. Mod. Phys. 53 (1981) 367;\l
       J. W. CRONIN: ``$CP$ symmetry violation - the search for its
             origin'', Rev. Mod. Phys. 53 (1981) 373.

\bibitem{Argus}
       H. ALBRECHT {\em et al.} (ARGUS collaboration): ``Observation of
             $B^0-\ol{B^0}$ mixing'', Phys. Lett. B 245 (1987), 245.

\bibitem{Nir}
       see for example:\l
       Y. NIR: ``$CP$ Violation'', Lectures given at 20th Annual SLAC Summer
            Institute on Particle Physics: The Third Family and the Physics
             of Flavor (School: Jul 13-24, Topical Conference: Jul 22-24,
             Symposium on Tau Physics: Jul 24), Stanford, CA, 13-24 Jul 1992.
             Published in SLAC Summer Inst.1992:81-136 (QCD161:S76:1992).

\bibitem{AdlerBellJackiw}
        S. L. ADLER: ``Axial-Vector Vertex in Spinor Electrodynamics'', 
                                        Phys. Rev. 177 (1969) 2426;\l
        J. S. BELL and R. JACKIW: Nuovo Cimento 60 (1969) 47;\l
        W. A. BARDEEN: ``Anomalous Ward Identities in Spinor Field
                                  Theories'', Phys. Rev. 184 (1969) 1848.

\bibitem{Ramond}
        P. RAMOND: ``Field Theory; a Modern Primer'', p. 226;
                 Frontiers in Physics,
                 Lecture Notes Series 51 (Benjamin/Cummings 1981), p. 226.

\bibitem{GellMannRamondSlansky} M. GELL-MANN, P. RAMOND and R. SLANSKY:
     ``Supergravity''  p. 315 (P. van Nieuwenhuizen and D.Z. Freedman eds,
       North Holland, Amsterdam 1979).

\bibitem{ChengLi} T.P. CHENG and Ling Fong LI: ``Neutrino masses, mixings,
                and oscillations in $SU(2) \times U(1)$ models of electroweak
                interactions'', Phys.Rev. D 22 (1980) 2860.

\bibitem{NambuJonaLasinio} Y. NAMBU and G. JONA-LASINIO: ``Dynamical Model
               of Elementary Particles Based on an Analogy with
               Superconductivity'' Phys. Rev. 122 (1961) 345; ibidem, 124
               (1961) 246.

\bibitem{ApplequistCarazzone} T. APPLEQUIST and J. CARRAZONE: ``Infrared
           singularities and massive fields'', Phys.Rev. D 11 (1975) 2856.

\bibitem{LeeWeinberg}
       T. D. LEE: ``A Theory of Spontaneous $T$ Violation'',
                      Phys. Rev. D 8 (1973) 1226;\l
       S. WEINBERG: ``Gauge Theory of $CP$ Nonconservation'', Phys. Rev.
                      Lett. 37 (1976) 657.

\bibitem{Petcov}
       S. M. BILENKI \& S. T. PETCOV: ``Massive neutrinos and neutrino
                            oscillations'', Rev. Mod. Phys. 59 (1987) 671.

\end{thebibliography}
\end{document}